\documentclass[twocolumn]{aastex63}
\graphicspath{{Figure/}}
\usepackage{graphicx}
\usepackage{subfigure}

\usepackage{float}

\def\h       {\ifmmode{^{\rm h}}\else{$^{\rm h}$}\fi}
\def\m       {\ifmmode{^{\rm m}}\else{$^{\rm m}$}\fi}
\def\s       {\ifmmode{^{\rm s}}\else{$^{\rm s}$}\fi}
\def\deg     {\ifmmode{^{\circ}}\else{$^{\circ}$}\fi}

\def\decdeg  {\ifmmode{{\rlap.}^{\circ}} \else ${\rlap.}^{\circ}$\fi}
\def\decs    {\ifmmode{{\rlap.}^{\rm s}} \else ${\rlap.}^{\rm s}$\fi}
\def\decas   {\ifmmode{{\rlap.}{''}}\else{${\rlap.}{''}$}\fi}

\def\K     {$K$}

\usepackage{color}
\definecolor{mygreen}{RGB}{44,85,17}                
\definecolor{myblue}{RGB}{34,31,217}                 
\definecolor{mybrown}{RGB}{194,164,113}          
\definecolor{myred}{RGB}{255,66,56}                 
\definecolor{mygrey}{RGB}{211,211,211}                 

\received{TBD}
\revised{TBD}
\accepted{TBD}
\submitjournal{AJ}

\shorttitle{Orbital period of colliding-wind binary WR 146}
\shortauthors{Wen et al.}

\begin{document}

\title{The orbital period of the long-period and colliding-wind binary WR 146 from radio interferometry of the shock cone
\footnote{Released on January 18th, 2025}}

  \correspondingauthor{Bo Zhang}
  \email{zb@shao.ac.cn}
  \correspondingauthor{Lang Cui}
  \email{cuilang@xao.ac.cn}

\author[0009-0008-1361-4825]{Shiming Wen}
    
\affiliation{Shanghai Astronomical Observatory, Chinese Academy of Sciences,
    80 Nandan Rd. 
    Xuhui, Shanghai, 200030, P.R.China}
    
\affiliation{Xinjiang Astronomical Observatory, Chinese Academy of Sciences, 
    150 Science 1-Street
    Urumqi, Xinjiang, 830011, P.R.China}

\affiliation{School of Astronomy and Space Science, University of Chinese Academy of Sciences, 
    No.19(A) Yuquan Rd.  
    Shijingshan, Beijing, 100049, P.R.China}

\author{Bo Zhang}
\affiliation{Shanghai Astronomical Observatory, Chinese Academy of Sciences, 
    80 Nandan Rd. 
    Xuhui, Shanghai, 200030, P.R.China}

\author[0000-0003-2953-6442]{Shuangjing Xu} 
\affiliation{Korea Astronomy and Space Science Institute,
    776 Daedeok-daero
    Yuseong-gu, Daejeon 34055, Republic of Korea}
    
\affiliation{Shanghai Astronomical Observatory, Chinese Academy of Sciences, 
    80 Nandan Rd. 
    Xuhui, Shanghai, 200030, P.R.China}
    
\author{Yan Sun}
\affiliation{Shanghai Astronomical Observatory, Chinese Academy of Sciences,
    80 Nandan Rd. 
    Xuhui, Shanghai, 200030, P.R.China}
    
\author[0000-0001-7573-0145]{Xiaofeng Mai}
\affiliation{Shanghai Astronomical Observatory, Chinese Academy of Sciences, 
    80 Nandan Rd. 
    Xuhui, Shanghai, 200030, P.R.China}

\affiliation{School of Astronomy and Space Science, University of Chinese Academy of Sciences, 
    No.19(A) Yuquan Rd.  
    Shijingshan, Beijing, 100049, P.R.China}
    
\author[0000-0002-9768-2700]{Jingdong Zhang}
\affiliation{Shanghai Astronomical Observatory, Chinese Academy of Sciences, 
    80 Nandan Rd. 
    Xuhui, Shanghai, 200030, P.R.China}

\affiliation{School of Astronomy and Space Science, University of Chinese Academy of Sciences, 
    No.19(A) Yuquan Rd.  
    Shijingshan, Beijing, 100049, P.R.China}
    
\author[0000-0003-0721-5509]{Lang Cui}
\affiliation{Xinjiang Astronomical Observatory, Chinese Academy of Sciences,
    150 Science 1-Street
    Urumqi, Xinjiang, 830011, P.R.China}
    
\affiliation{Key Laboratory of Radio Astronomy and Technology (Chinese Academy of Sciences),
    A20 Datun Road
    Chaoyang District, Beijing, 100101, P.R.China}
    
\affiliation{Xinjiang Key Laboratory of Radio Astrophysics, 
    150 Science 1-Street, 
    Urumqi, Xinjiang, 830011, P.R.China}
    
\author[0000-0002-9093-6296]{Xiaofeng Li}  

\affiliation{School of Computer Science and Information Engineering,
    Changzhou Institute of Technology,
    Changzhou, Jiangsu 213032, P.R.China}
    
\author[0000-0002-9684-3074]{Helge Todt}

\affiliation{Institut für Physik und Astronomie, Universität Potsdam, Karl-Liebknecht-Str. 24/25, D-14476 Potsdam, Germany}   
    
\author{Xi Yan}

\affiliation{Xinjiang Astronomical Observatory, Chinese Academy of Sciences, 
    150 Science 1-Street
    Urumqi, Xinjiang, 830011, P.R.China}

\author{Pengfei Jiang}

\affiliation{Xinjiang Astronomical Observatory, Chinese Academy of Sciences, 
    150 Science 1-Street
    Urumqi, Xinjiang, 830011, P.R.China}
    
\begin{abstract}

We report the first measurement of the orbital period of a long-period colliding-wind binary (CWB) system WR 146, derived by tracing the rotational morphology of its wind-colliding region (WCR) and the relative orientation of the two binary components. This result is based on our imaging observations using the Very Long Baseline Array (VLBA) and the European Very Long Baseline Interferometry (VLBI) Network (EVN), combined with archival data from VLBA, EVN, the Very Large Array (VLA), the enhanced Multi-Element Radio-Linked Interferometer Network (eMERLIN) arrays, and optical images from the Hubble Space Telescope (HST). We evaluated two methods for determining the binary's orbital period based on the images of the WCR: (I) fitting the shock cone of the WCR and (II) stacking images using the cross-correlation function. Using these techniques, we find orbital period estimates of 810$^{+120}_{-90}$ years from method I and 1120$^{+540}_{-270}$ years from method II, both of which support a long orbital period of approximately 1,000 years. Furthermore, we analyzed archival spectral data of WR 146 to estimate the stellar wind velocities of the binary components, finding no significant orbital phase lag between the binary orientation and the WCR rotation. We also estimate the range of the binary's mass using the currently measured parameters.

\end{abstract}


\section{Introduction} \label{sec:intro}

Colliding-wind binaries (CWBs) are massive binary systems distinguished by intense stellar wind activity. The powerful winds from their massive components collide, forming strong shocks, known as wind-colliding regions (WCRs) at the shock cones. In WCRs, particles are accelerated to relativistic velocities. Relativistic particles then interact with the binary's magnetic field, producing non-thermal synchrotron emission. The mechanism of particle acceleration in WCRs is widely accepted to be diffusive shock acceleration (DSA) \citep{2005ApJ...623..447D}. As a unique laboratory for investigating particle acceleration, WCRs can provide denser matter, higher energy, and stronger magnetic fields than the normal interstellar medium environment (e.g., supernova remnants or planetary nebulae), more suitable for investigating the mechanisms of particle acceleration \citep{2013A&A...558A..28D}. Moreover, studying wind-colliding activities aids in understanding the late-stage evolution of massive stars by estimating wind mass-loss rates and modeling the shock wave fronts of the colliding winds \citep{2010ASPC..422..166D}. Binaries with massive stars, especially those containing Wolf-Rayet (WR) stars with strong stellar wind outflows, are more likely to form WCRs that emit non-thermal emission due to the large quantities of charged particles. \citep{2013A&A...558A..28D,2017A&A...600A..47D,2020MNRAS.495.2205P}. 

cDue to significant extinction at optical band (e.g., dust scattering) \citep{2022arXiv220412354E}, CWBs are more commonly detected in the radio wavelengths. Additionally, because the angular size of the WCR region is only 10-20 mas, Very-Long Baseline Interferometry (VLBI) observation is a unique method to image WCRs with the highest angular resolution \citep{2015ICRC...34..509G}. Due to the relative low non-thermal radiation flux, only a few CWB systems with detectable WCRs have been studied in detail, such as WR 86, WR 140, WR 147 \citep{1998AJ....115.2047N, 2010ASPC..422..166D}, HD167971 and HD168112 \citep{2024A&A...682A.160D}. Among them, WR 140 is the only typical CWB with fully determined orbital parameters. \citet{1990MNRAS.243..662W} suggested a stellar separation of 2–30 AU and an orbital period of 8 years, then \citet{2011BSRSL..80..658D} also derived a similar period by tracing the rotation of WCR. This result have been refreshed as 2896.5$\pm$0.7 days measured by \citet{2011MNRAS.418....2F}, 2896.35$\pm$0.2 days by \citet{2011ApJ...742L...1M}, 2895.0$\pm$0.29 days by \citet{2021MNRAS.504.5221T}, with the periodic variation of radial velocity. It exhibits that tracing the WCR rotation is an effective approach and worth adopting for binary period measurement.

WR 146 is one of the brightest WR stars at radio wavelengths \citep{2000A&A...356..676S}. It has been identified as a WC6+O8 CWB system projected towards the heavily obscured Cyg OB2 region \citep{1998AJ....115.2047N,1991AJ....101.1408M}. Previous radio interferometry observations have revealed that its thermal emission is mainly from the stellar wind of the two stars: southern WR and northern O star~\citep{2000MNRAS.316..143D}. The bright bow-shaped non-thermal emission is mainly from the WCR, where the winds collide~\citep{2005mshe.work...81O}. Observations at 43 GHz with Very Large Array (VLA) combined with the radio telescope at Pie Town suggest that the apparent stellar separation is 152 $\pm$ 2 mas \citep{2005mshe.work...49D}, which is statistically consistent with that of 168 $\pm$ 31 mas measured by the Hubble Space Telescope (HST) \citep{1998AJ....115.2047N}. Gaia DR2 suggests that the binary is at a distance of 1.2 $^{+1.0}_{-0.4}$ kpc \citep{2018yCat.1345....0G, 2018AJ....156...58B}. And a total stellar mass of $\sim$50 M$_{\odot}$ implies its orbital period is greater than 350 yrs \citep{2010ASPC..422..166D}. 

The key parameters of CWB systems include the binary orbital parameters, with the orbital period being the most crucial. Due to the binary's long orbital period and the small angular separation of its components, conventional observational methods struggle to resolve the system's structure. As a result, directly observing the complete orbital motions of the binary components and accurately determining the orbital period remains challenge. Consequently, there are still rare CWB systems with well-determined orbital periods. However, we can still employ indirect methods, such as using VLBI technology to observe the morphology and positional changes of the WCR, to infer the relative positions of the binary components and further estimate the angular velocity of their orbital motion. CWB model described by \citet{1993ApJ...402..271E} and \citet{ 2003A&A...409..217D} demonstrates that the shock front formed by spherically symmetric binary stellar winds exhibits an axisymmetric shape, with the axis of symmetry aligned with the binary axis. Analysis combined observations in optical and radio bands by \citet{2005mshe.work...81O} also indicates that the symmetry axis of WR 146's WCR is also generally consistent with the binary axis. Therefore, the orientation of the WCR can be used here as a reference for determining the direction of the binary axis. This method have been succeed in short period CWB systems such as WR140 \citep{2011BSRSL..80..658D}. Therefore, we apply this method to observations of the WR 146 system to assess its feasibility for long-period CWBs, while also providing a constraint for the study of WR 146 and other similar systems.

In this paper, we present seven epochs of VLBI observations of WR 146 using the Very Long Baseline Array (VLBA) and the European VLBI Network (EVN), and describe our data reduction process in Section~\ref{sec:data}. Additionally, we have compiled a total of fifteen epochs of archival data from 1992 to 2016, obtained from the VLA, enhanced Multi-Element Radio-Linked Interferometer Network (eMERLIN), VLBA, and EVN archives. We also collected images from two epochs and a spectroscopic observation of WR 146 in the optical waveband, obtained with the HST. In Section~\ref{sec:orbit}, we present our methods and results based on all available data. In Section~\ref{sec:discussion}, we estimate the physical parameters of WR 146, including the orbital period and mass, and discuss some controversies of the WR 146 system. Finally, we present our summary in Section~\ref{sec:sum}.

\section{Data} 
\label{sec:data}   

The position angle of the binary components changes as they orbit each other, as indicated by variations in the relative orientation of the components and the morphology of the WCR. By tracking these changes, we can monitor the orbital motion of the system. To determine the position angles, we compiled multiple epochs of archival data from the VLA, MERLIN, EVN, VLBA, and HST, as listed in Tables~\ref{tab:VLAHST} and \ref{tab:radio}. We calibrated and imaged 22 epochs of radio interferometric data using the Astronomical Image Processing System (AIPS)\footnote{\url{http://www.aips.nrao.edu/index.shtml}} and measured the two epochs of HST images by DS9 \footnote{\url{https://cfa.harvard.edu/saoimageds9/home}}. Given the different emission characteristics, we divided the data into two groups: one for the binary components and the other for the WCR.

\subsection{WCR}

Observations show that the radio emission of CWB systems primarily includes thermal radiation from the binary stars and non-thermal synchrotron radiation from the WCR \citep{2010ASPC..422..166D}.. Non-thermal synchrotron radiation dominates in the frequency range below 20 GHz, peaking at $\sim$0.5 GHz \citep{2003A&A...409..217D}. We chose to image the WCR at the C band (5 GHz), as it provides the optimal combination of sensitivity and resolution for our study. Table~\ref{tab:radio} lists the radio interferometer observations using EVN, VLBA and MERLIN, which are capable of resolving the non-point-like 
structure of the WCR. 

\begin{figure*}[htbp]
  \centering

    \subfigure[]{
        \begin{minipage}[t]{0.3\linewidth}
            \centering
            \includegraphics[width=6cm]{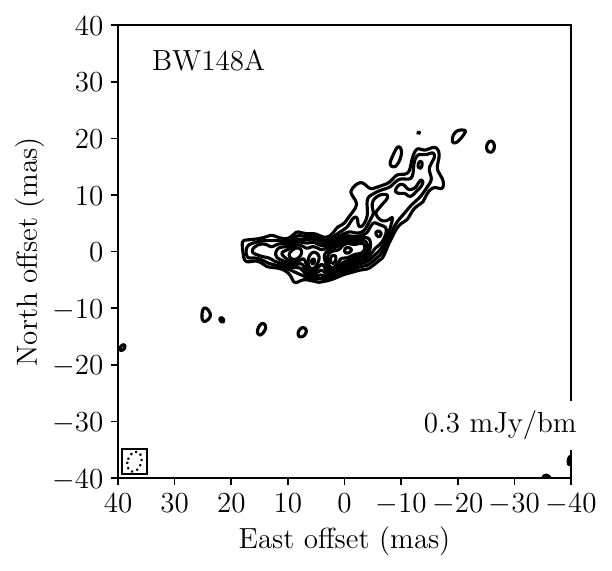}
        \end{minipage}%
    }%
    \subfigure[]{
        \begin{minipage}[t]{0.3\linewidth}
            \centering
            \includegraphics[width=6cm]{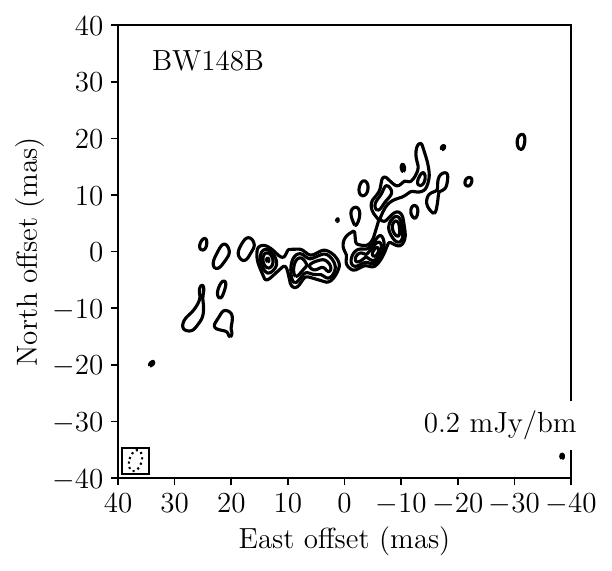}
        \end{minipage}%
    }%
    \subfigure[]{
        \begin{minipage}[t]{0.3\linewidth}
            \centering
            \includegraphics[width=6cm]{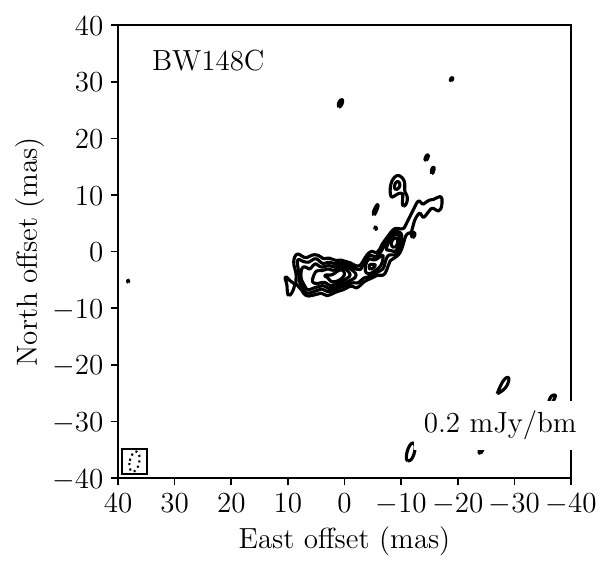}
        \end{minipage}%
    }%
    \quad
    \subfigure[]{
        \begin{minipage}[t]{0.3\linewidth}
            \centering
            \includegraphics[width=6cm]{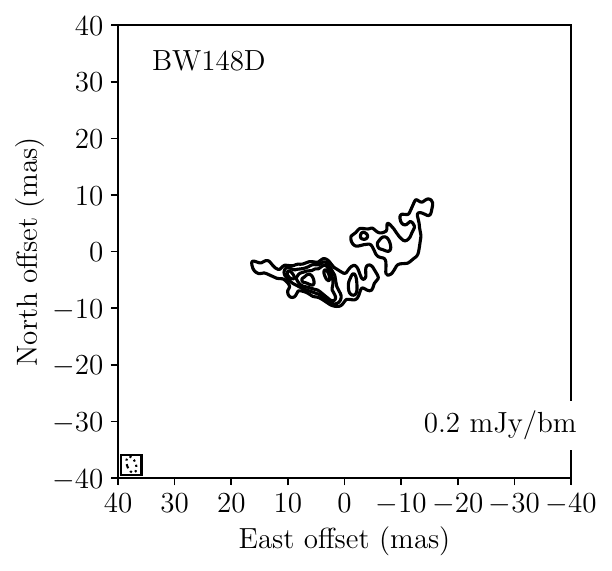}
        \end{minipage}%
    }%
    \subfigure[]{
        \begin{minipage}[t]{0.3\linewidth}
            \centering
            \includegraphics[width=6cm]{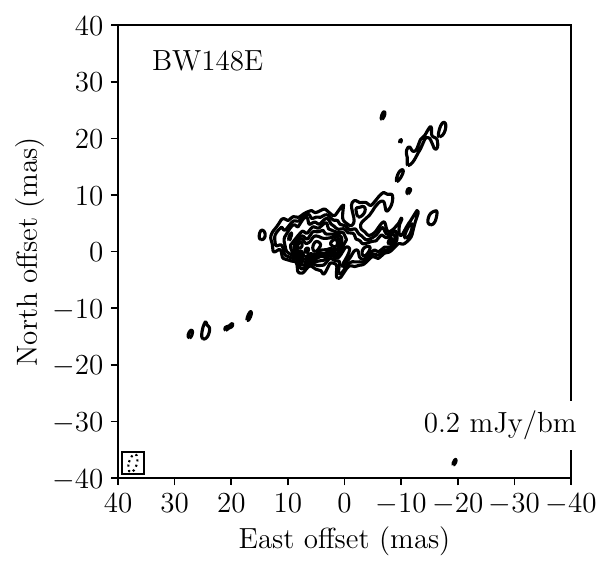}
        \end{minipage}%
    }%
    \subfigure[]{
        \begin{minipage}[t]{0.3\linewidth}
            \centering
            \includegraphics[width=6cm]{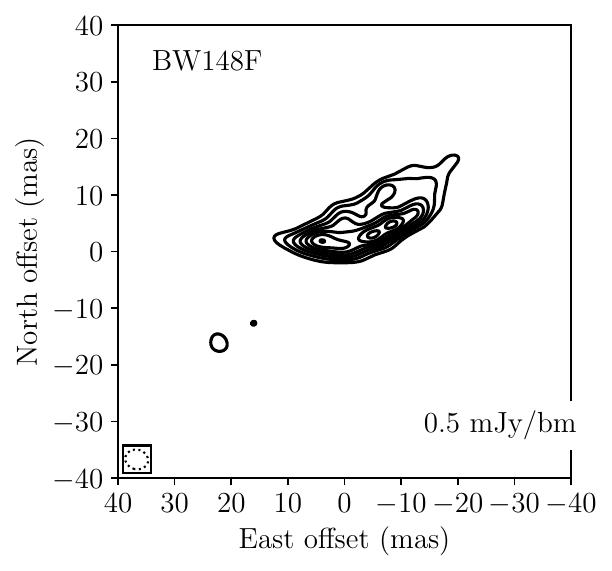}
        \end{minipage}%
    }%
    \quad
        \subfigure[]{
        \begin{minipage}[t]{0.3\linewidth}
            \centering
            \includegraphics[width=6cm]{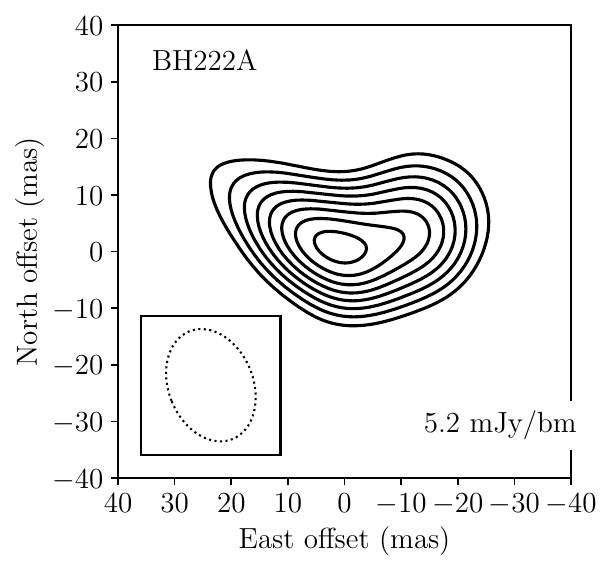}
        \end{minipage}%
    }%
    \subfigure[]{
        \begin{minipage}[t]{0.3\linewidth}
            \centering
            \includegraphics[width=6cm]{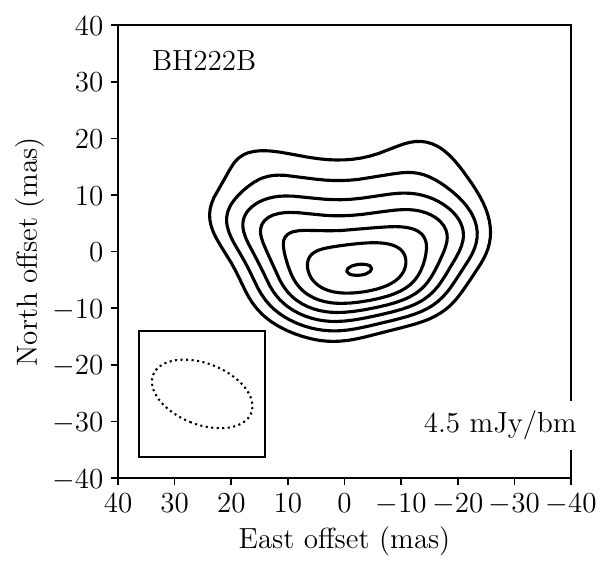}
        \end{minipage}%
    }%
    \subfigure[]{
        \begin{minipage}[t]{0.3\linewidth}
            \centering
            \includegraphics[width=6cm]{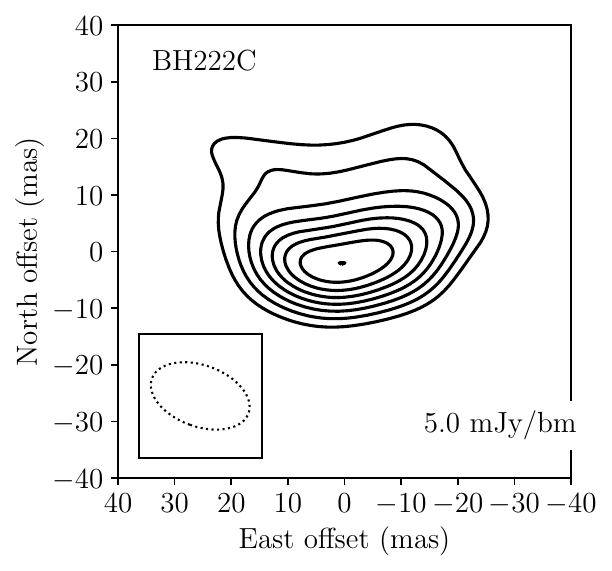}
        \end{minipage}%
    }%
    \quad

    \caption{
    EVN and VLBA images of WCR of WR 146 at C band (5 GHz), part 1. \textit{(a)$\sim$(f): } VLBA images of BW148 series with resolution $\sim$3 mas. The rms is $\sim$0.3 mJy~bm$^{-1}$, and contour level is (3,4,5,6,7,8,9,10) $\times$ 0.03 mJy\,bm$^{-1}$. \textit{(g)$\sim$(i):} VLBA image of BH222 series A$\sim$C with restoring beam size of $\sim$20 mas, and contour level is (3,4,5,6,7,8,9,10) $\times$ 0.5 mJy\,bm$^{-1}$.}
    \label{fig:WR 146(1)}  
\end{figure*}

\begin{figure*}[htbp]
  \centering
        \subfigure[]{
        \begin{minipage}[t]{0.3\linewidth}
            \centering
            \includegraphics[width=6cm]{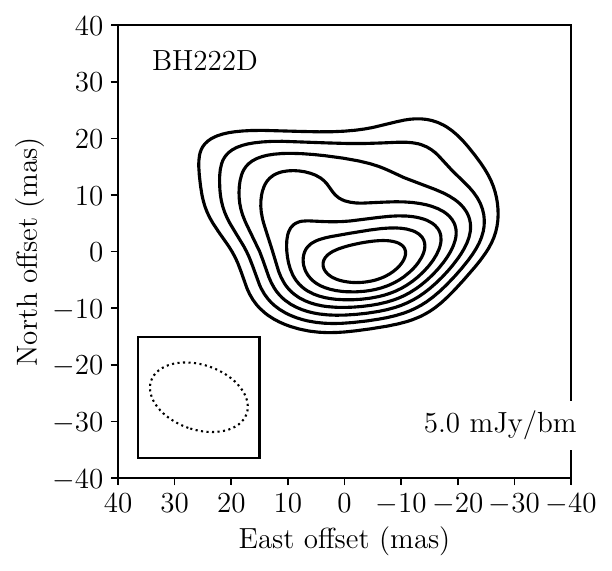}
        \end{minipage}%
    }%
    \subfigure[]{
        \begin{minipage}[t]{0.3\linewidth}
            \centering
            \includegraphics[width=6cm]{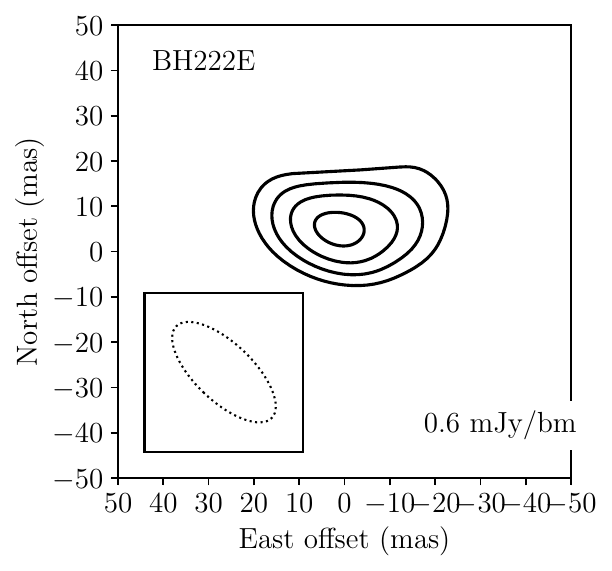}
        \end{minipage}%
    }%
    \subfigure[]{
        \begin{minipage}[t]{0.3\linewidth}
            \centering
            \includegraphics[width=6cm]{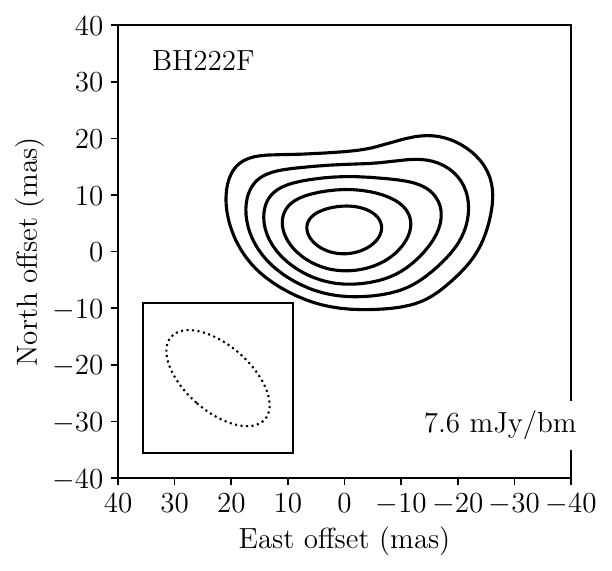}
        \end{minipage}%
    }%
    \quad
        \subfigure[]{
        \begin{minipage}[t]{0.3\linewidth}
            \centering
            \includegraphics[width=6cm]{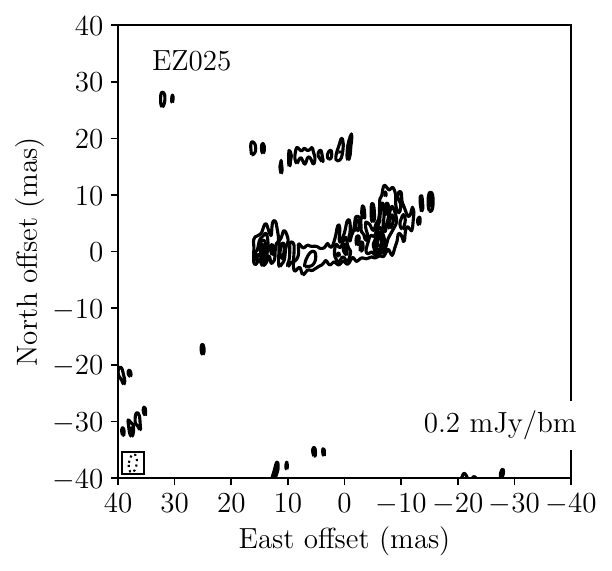}
        \end{minipage}%
    }%
    \subfigure[]{
        \begin{minipage}[t]{0.3\linewidth}
            \centering
            \includegraphics[width=6cm]{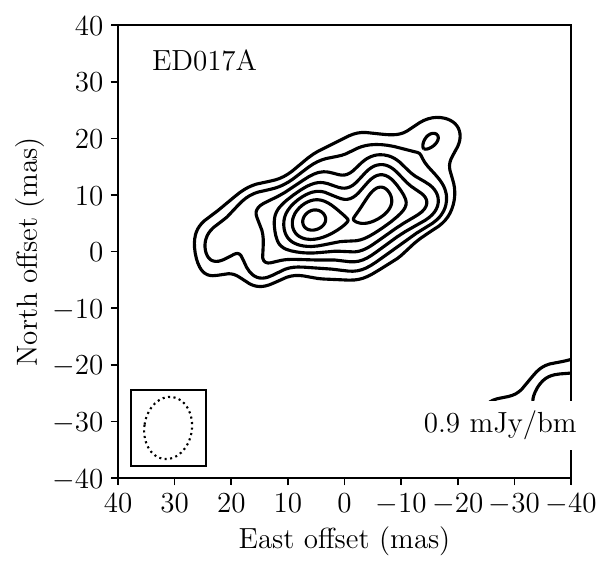}
        \end{minipage}%
    }%
    \subfigure[]{
        \begin{minipage}[t]{0.3\linewidth}
            \centering
            \includegraphics[width=6cm]{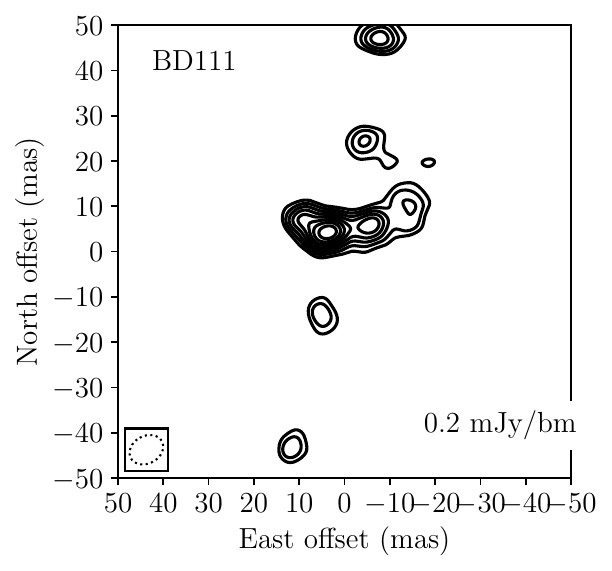}
        \end{minipage}%
    }%

    \caption{
EVN and VLBA images of WCR of WR 146 at C band (5 GHz), part 2. (a)$\sim$(c):  VLBA image of BH222 series D$\sim$F with restoring beam size of $\sim$20 mas, and contour level is (3,4,5,6,7,8,9,10) $\times$ 0.5 mJy\,bm$^{-1}$ for (a), (3,4,5) $\times$ 0.1 mJy\,bm$^{-1}$ for (b), (3,4,5,6,7) $\times$ 1 mJy\,bm$^{-1}$ for (c). (d): EVN image EZ025 with restoring beam size of 5$\times$4 mas$^2$. The rms is 0.2 mJy~bm$^{-1}$ and contour level is (3,4,5) $\times$ 0.04 mJy\,beam$^{-1}$. (e): EVN image ED017a with restoring beam size of 11$\times$8 mas$^2$, and contour level is (3,4,5,6,7,8,9) $\times$ 0.1 mJy\,bm$^{-1}$. (f): VLBA image BD111 with restoring beam size of 8$\times$6 mas$^2$, and contour level is (3,4,5,6,7,8,9,10) $\times$ 0.02 mJy\,bm$^{-1}$. }
    \label{fig:WR 146(2)}    
\end{figure*}

\subsubsection{EVN and VLBA Data}

To investigate the WCR evolution over a longer time baseline, we carried out a single-epoch 5~GHz observation of WR 146 employing EVN on May 30th, 2016 (program code: EZ025) to map the WCR and compare the image with those from previous VLBI observations. The observation was conducted in phase-referencing mode because the WCR of WR 146 is a weak target, so this mode was used to improve the signal-to-noise ratio (SNR) with longer integration time and hence detect the weak emission. There were nine participating telescopes: Jodrell Bank (Jb-2), Westerbork (Wb), Effelsberg (Ef), Medicina (Mc), Noto (Nt), Onsala-85 (On-85), Torun (Tr), Yebes (Ys) and Irbene (Ir). The observation was performed with a data rate of 1 Gbps (two circular polarizations and eight contiguous 16 MHz subbands with 2-bit sampling). 

The observation sequences included WR 146, along with calibrators J2015+3710 and J2007+4029, with J2007+4029 preferred as the phase calibrator, switching every 2 minutes between the target and quasars. Three sources: J1955+5131, J2002+4725 and J2212+2355 were observed at the beginning, middle and end of the observation for bandpass calibration. The total observation time was 16 hours, with 8 hours spent on WR 146.

Subsequently, we proposed a series of six observations at 5~GHz by VLBA, spanning from April 27th, 2023 to June 16th, 2024 (program code: BW148). Like the EVN observation, we also conducted phase-referencing observations. A total of  ten antennas were involved in all or part of these observations: Brewster(BR), Kitt Peak(KP), Los Alamos(LA), North Liberty(NL), Pie Town(PT), Saint Croix(SC) were used in all observations, while Fort Davis(FD), Hancock(HN), Mauna Kea(MK), Ovens Valley(OV) participated in some of the observations. We adopted a recording rate of 4 Gbps (including 8 dual-polarized frequency bands of 128 MHz bandwidth with 256 channels and 2-bit sampling) for adequate imaging sensitivity. We preferred J1058+8114 for bandpass calibration, and four sources, J2007+4029, J2032+4057, J2041+4527 and J2050+3620, for phase calibration. Each observation took 8 hours for each observation, with 4 hours spent on WR 146. All processed data present collectively in Table~\ref{tab:radio}.

Figure~\ref{fig:WR 146(1)} and \ref{fig:WR 146(2)} displays the images of the WCR obtained from both EVN and VLBA observations. The WCR was further resolved into a bow-shaped structure with a clear boundary, oriented toward the O8-type star. At relative lower frequency wavelengths ($\leq$ 20 GHz), the radio emission from this system is primarily synchrotron radiation originating from the WCR \citep{2010ASPC..422..166D}, which is symmetric about the line connecting the two stars, with the WCR pointing toward the O8-type star \citep{1993ApJ...402..271E}. Therefore, by mapping the non-thermal emission with VLBI observations, we can monitor the movement of the WCR and trace its rotation across different epochs, providing a potential method for tracking the binary's orbital motion \citep{2005ApJ...623..447D}.
 
\subsubsection{VLBI archive data}

We also retrieved VLBA and EVN archive data, collecting a total of 8 epochs from three series: VLBA's BD111, BH222, and EVN's ED017a. We also processed these data for imaging, and the resulting images are displayed in Figure~\ref{fig:WR 146(1)} and~\ref{fig:WR 146(2)}. The datasets were calibrated using the standard procedure for VLBI phase-referencing in AIPS and then imaged using AIPS task IMAGR, following a procedure similar to that used for BW148.

\subsubsection{eMERLIN Data}

\begin{figure*}[htbp]
    \centering
    \includegraphics[angle=0,scale=0.2]{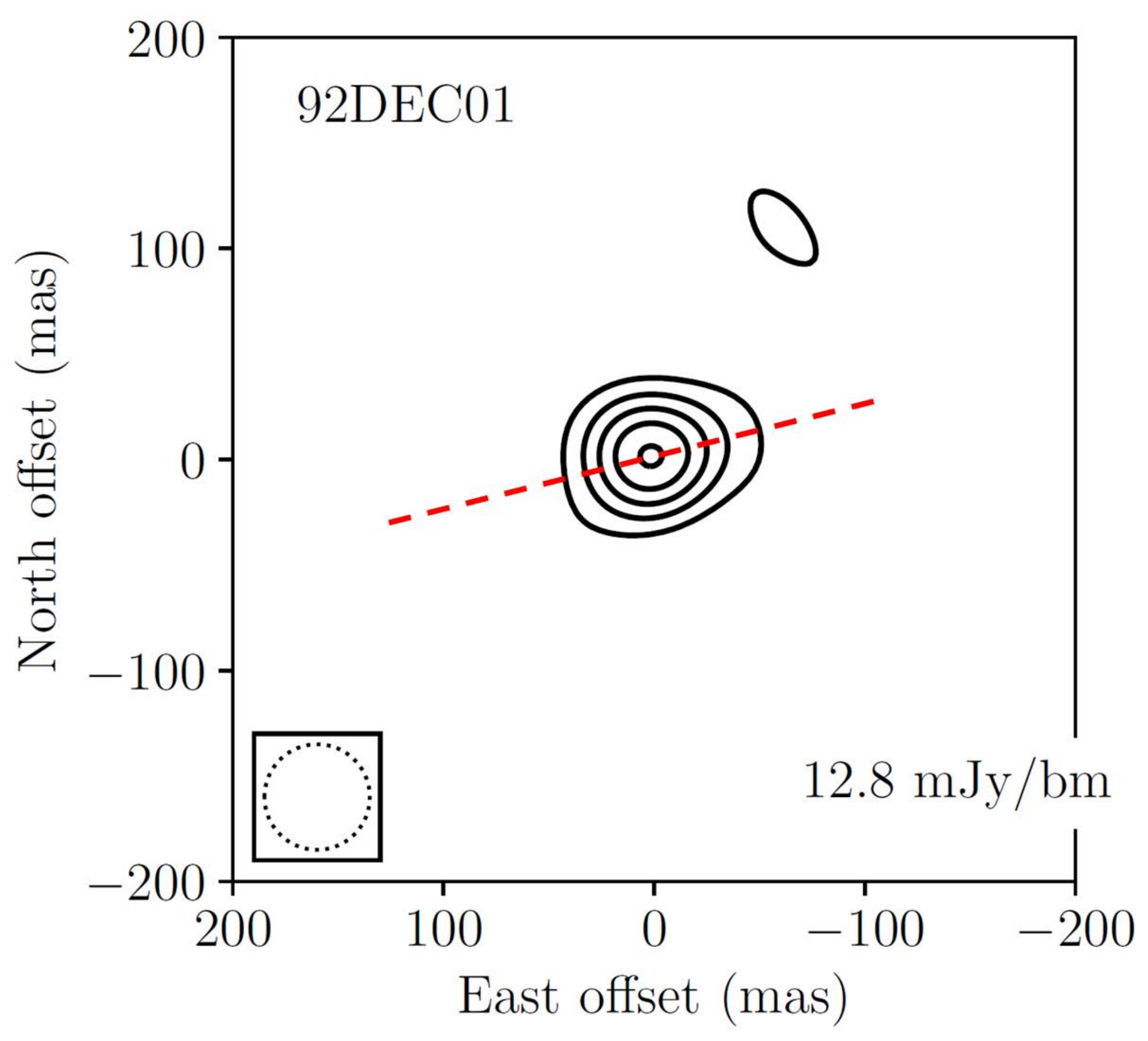}
    \includegraphics[angle=0,scale=0.2]{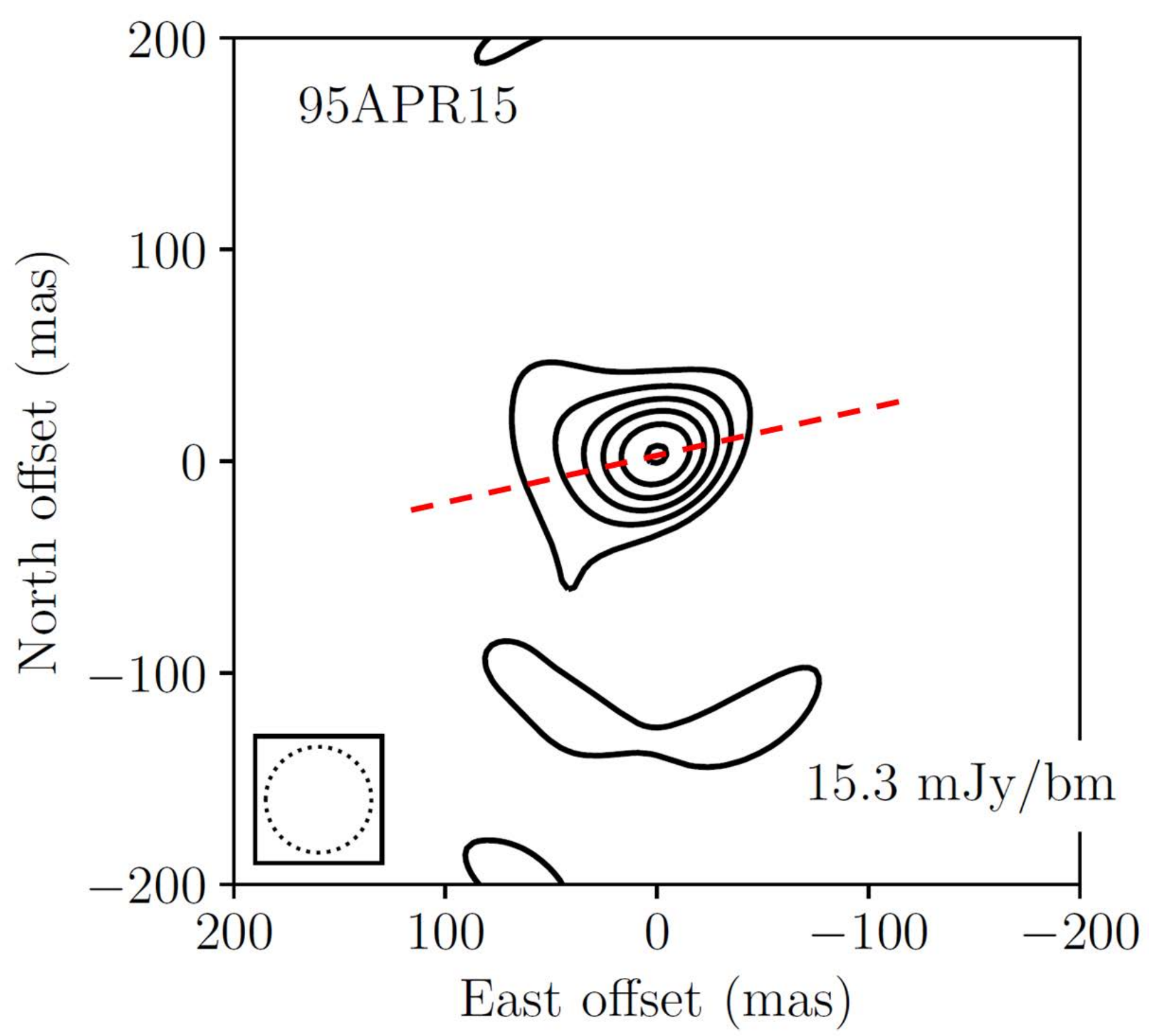}
    \includegraphics[angle=0,scale=0.2]{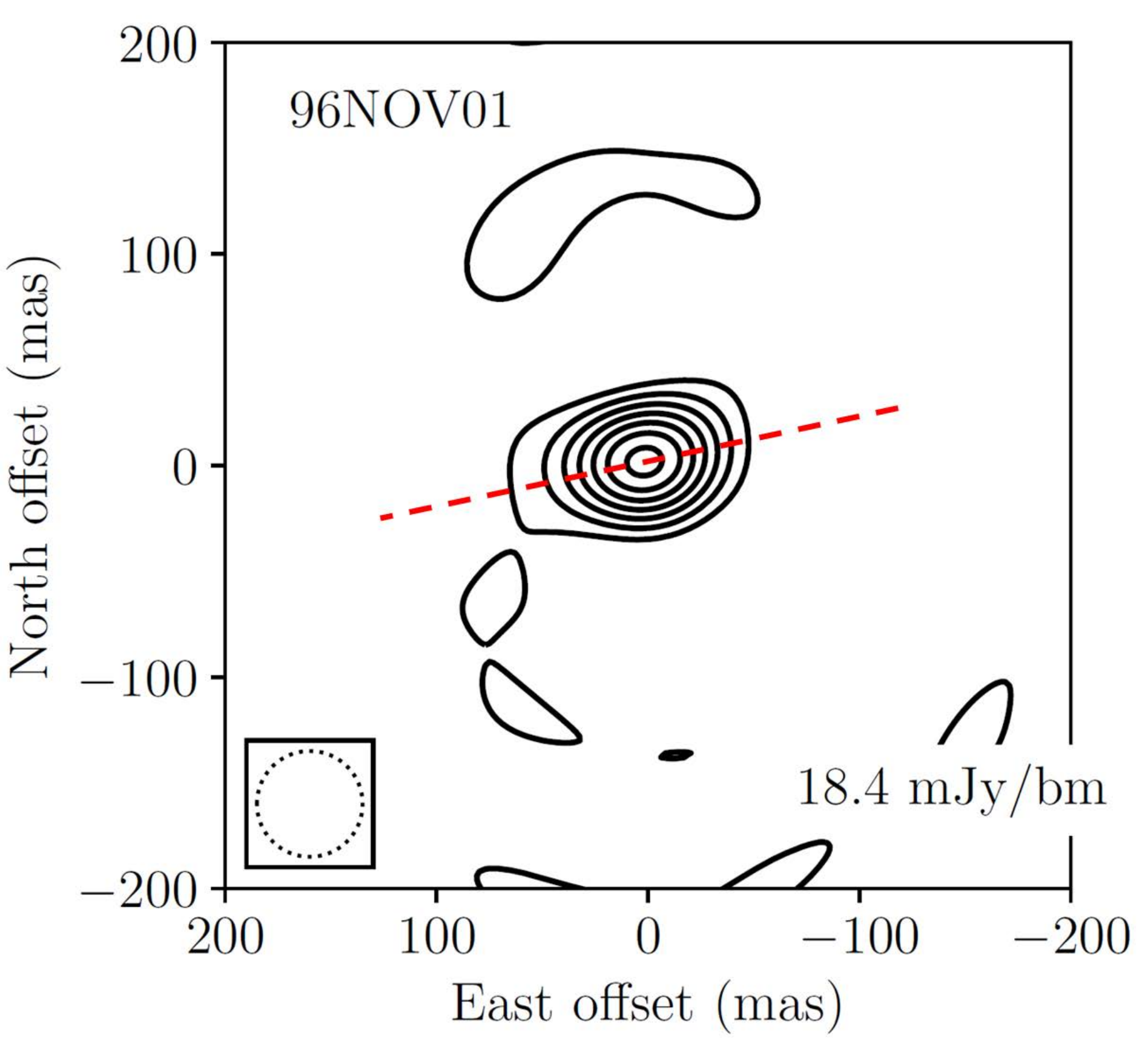}
    \includegraphics[angle=0,scale=0.2]{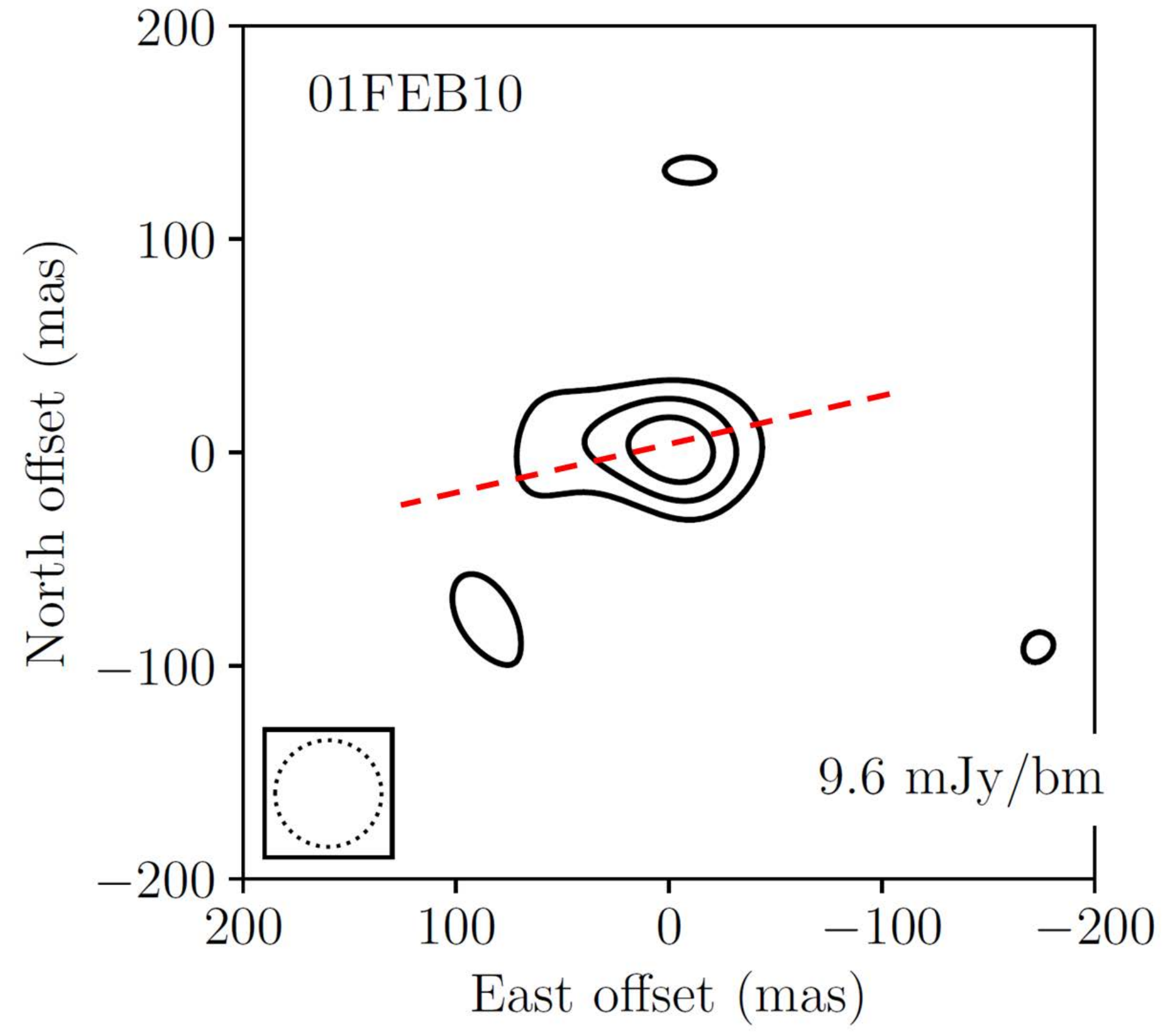}
    \caption{MERLIN images at 5 GHz of WR146 were obtained at the following epochs 1992-DEC-01, 1999-APR-15, 1996-NOV-01 and 2001-FEB-10. 
    The restoring beam size are 50 mas, with contour levels at (1, 2, 3, 4, 5, 6) $\times$ 1$\sigma$ (2.5mJy). The restoring beams are marked in the lower left corner of each panel. 
    The long axis of the oval structure, indicated by red dashed lines in each panel, is used to trace the orientation of the WCR,  which is assumed to be perpendicular to the the line connecting the two stars.
    \label{fig:merlin}}
\end{figure*}

There are four epochs well-calibrated archive data collected from the MERLIN Archive ~\footnote{\url{https://www.e-merlin.ac.uk/archive/}}. To facilitate comparison of the morphology across different epochs, we applied a uniform restoring beam of 50 mas × 50 mas, using a circular beam for all images.
Imaging was performed using the AIPS task IMAGR. Due to the limited angular resolution of eMERLIN, the WCR could only be resolved as an elliptical structure, as shown in Figure~\ref{fig:merlin}, rather than the clearly bow-shaped structure observed in the higher-resolution VLBI data.

\subsection{Binary components}

\begin{figure*}[htbp]
    \centering
    \includegraphics[width=1\textwidth]{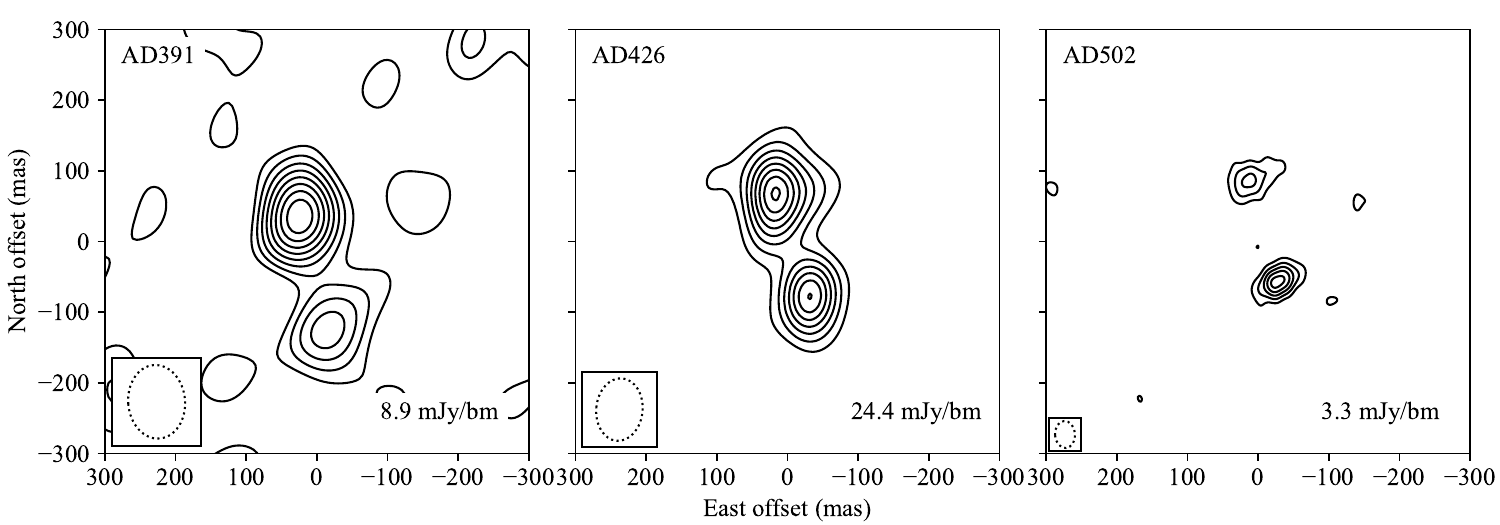} 
    \caption{Images of binary components in WR~146 from VLA observations in A configuration at K and Q bands. 
    For each panel, the program code and peak brightness are labeled in the top left corner and bottom right corner, respectively.  
    The restoring beam is shown in the bottom left corner. Contour levels for all panels start from (3,4,5,6,7,8,9)$\times$1$\sigma$ (0.8, 2.7 and 0.5 mJy~bm$^{-1}$ from left to right, respectively.). 
    The details of the observations are listed in Table~\ref{tab:VLAHST}. 
    \label{fig:VLA}
    }            
\end{figure*}

In the radio K and Q bands, the radiation from WR 146 is primarily dominated by thermal emission from the binary stars \citep{2003A&A...409..217D}. VLA observations in these bands can map this thermal emission from both stars, allowing for the resolution of the individual members of WR 146. Similarly, optical observations with the HST can also resolve and image the binary stars. Consequently, the position angle of the binary can be determined accurately.

\subsubsection{VLA data}

We retrieved 9 epochs of VLA data for WR 146 from the National Radio Astronomy Observatory (NRAO) Archive ~\footnote{\url{https://data.nrao.edu/}} and processed the data using standard imaging procedures in AIPS. Of these, 3 epochs (program codes: AD391, AD426, and AD502) were observed in the A-array configuration (baseline length $\sim$ 20 km) at K/Q band, providing sufficient angular resolution to determine the relative position of the binary stars. The remaining epochs were observed with B, C, or D-array configurations, which have shorter baselines and/or lower frequencies, resulting in lower angular resolution and an unresolved binary. Figure~\ref{fig:VLA} displays the VLA images of the resolved binary.

\subsubsection{HST data}

\begin{figure}[H]
    \centering

    \includegraphics[width=0.206\textwidth]{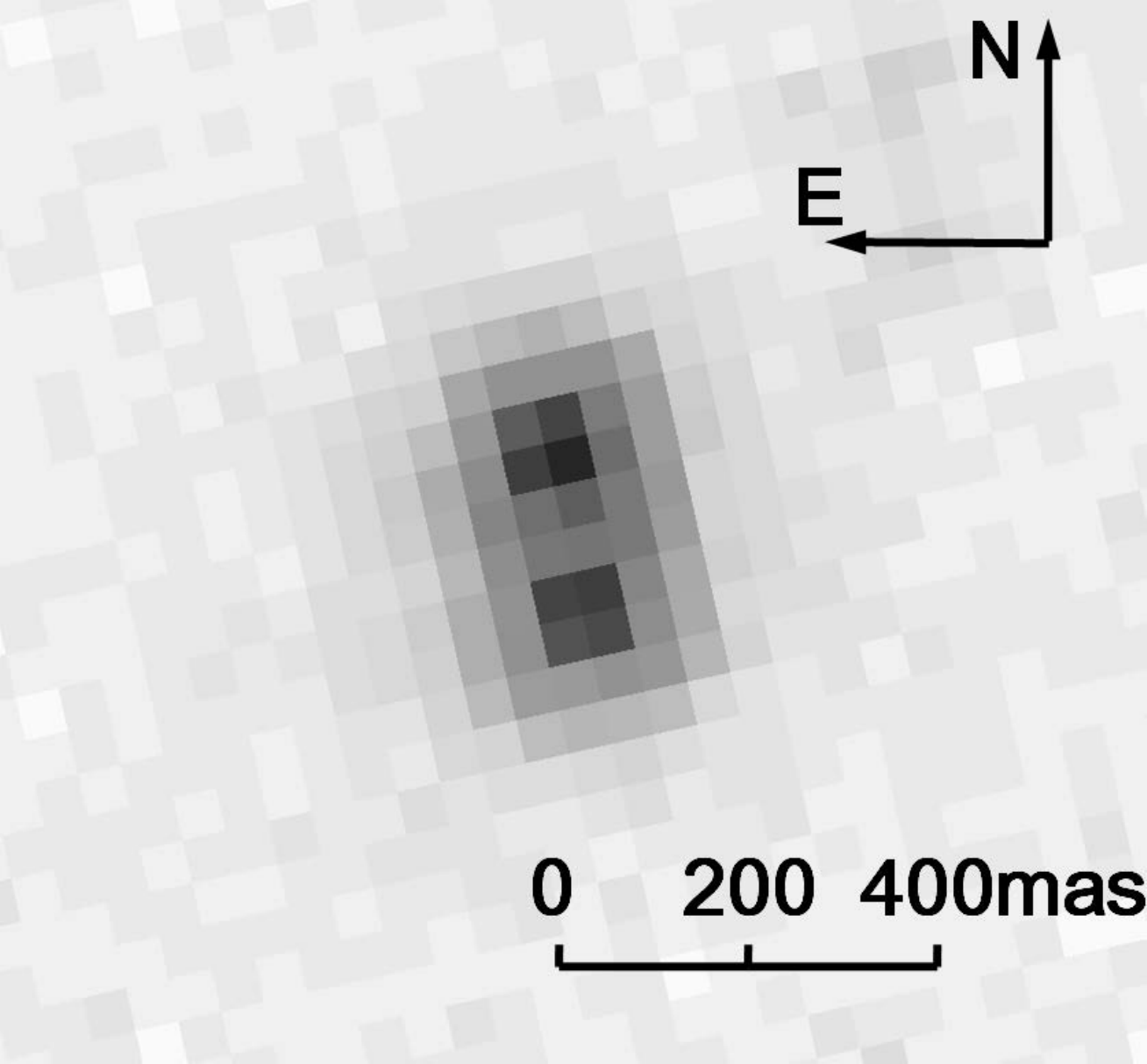} 
    \includegraphics[width=0.2\textwidth]{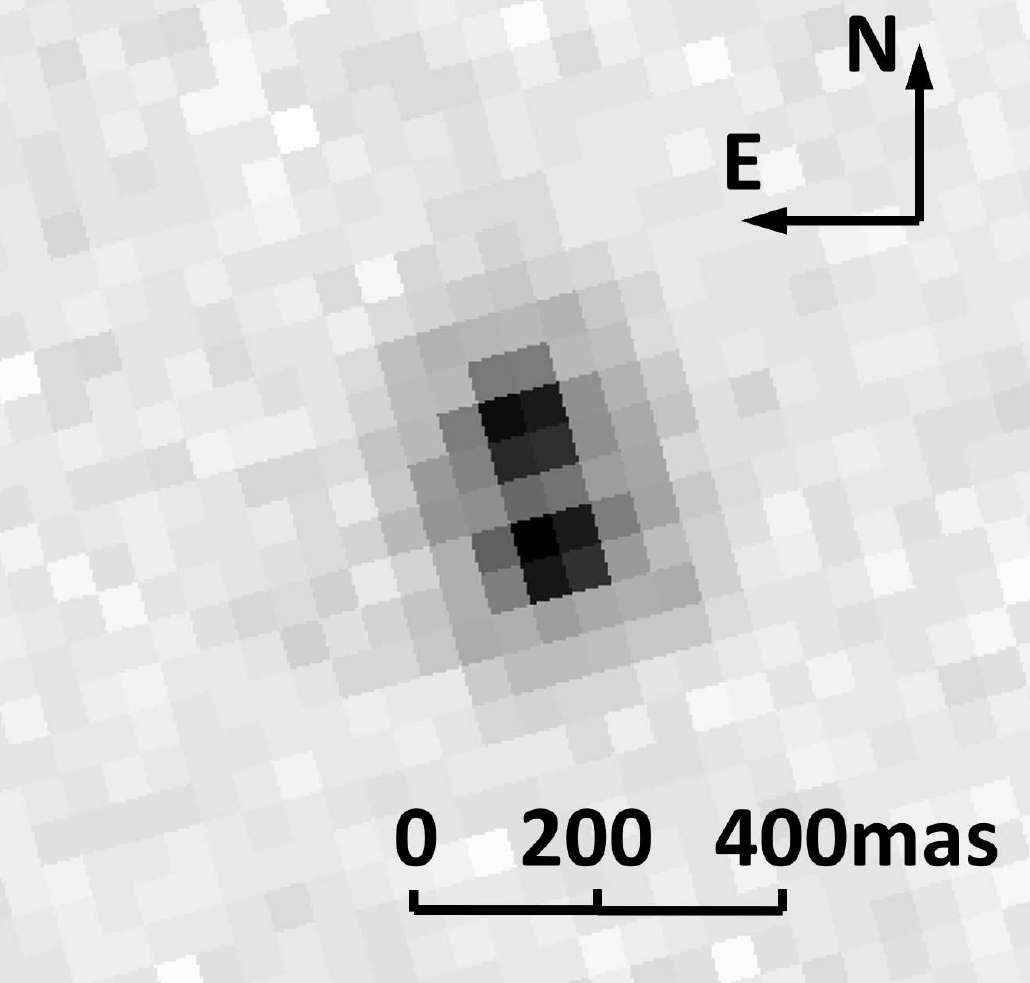} 
    \caption{Gray-scale photo of WR~146 from the HST observations with the Wide Field and Planetary Camera (WFPC) in 1996 (left panel) and with the Space Telescope Imaging Spectrograph (STIS) in 2000 (right panel). The photos are in same scale but with different pixel size. 
    }
    \label{fig:HST}
\end{figure}

The HST archival images of WR 146 were obtained from the Canadian Astronomy Data Centre (CADC) \footnote{\url{https://www.cadc-ccda.hia-iha.nrc-cnrc.gc.ca/en/}}, with two epochs observed at 5550\AA, corresponding to an angular resolution of approximately 50 mas. Details of the observations are provided in Table \ref{tab:VLAHST}. The images were processed through the standard HST pipeline \citep[HST Data Handbook,][] {1995stid.book....2L} as described in~\citet{1998AJ....115.2047N} and \citet{2001AJ....122.3407L}. Figure~\ref{fig:HST} shows a well-resolved HST image of the two individual members of WR 146. 
 
\section{Results} 
\label{sec:orbit} 

\subsection{Position angle of binary components and WCR in WR 146}
\label{sec:positionangle}

For visual binaries, sufficient data on the relative positions of the binary components can be utilized to determine their orbital motion.
Unfortunately, for WR146, the orbital period is likely much longer than the observed time span.
While limited data may not allow for precise orbit determination, we can still infer the orbital period from the available information, particularly since we have added new epochs to extend the observation time span. The change in the position angle of the binary reflects its orbital motion and can be measured directly.
In addition, as shown in Figures \ref{fig:WR 146(1)} and \ref{fig:WR 146(2)}, the WCR exhibits a persistent bow-shaped structure across all epochs. The orientation of the WCR, with its symmetric axis pointing toward the binary, aligns with the position angle of the binary due to the stellar wind collision, as depicted in Figure \ref{fig:td}. This alignment can be used to indicate the orientation of the binary components. Therefore, the orbital period can be estimated by measuring the changes in both the position angle of the binary and the orientation of the WCR over time.

\begin{figure}
    \centering    
    \includegraphics[width=0.48\textwidth]{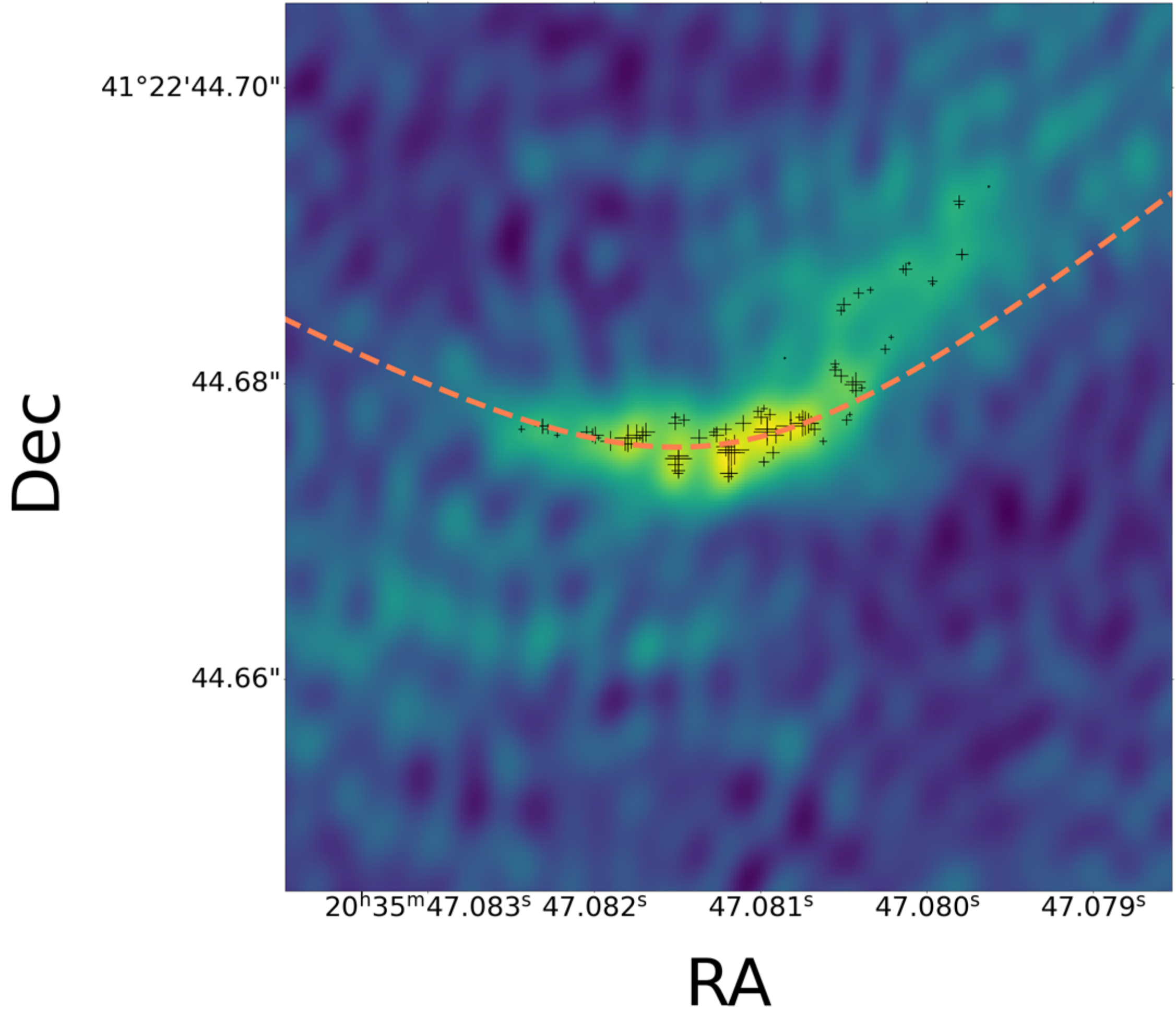} 
    \includegraphics[width=0.48\textwidth]{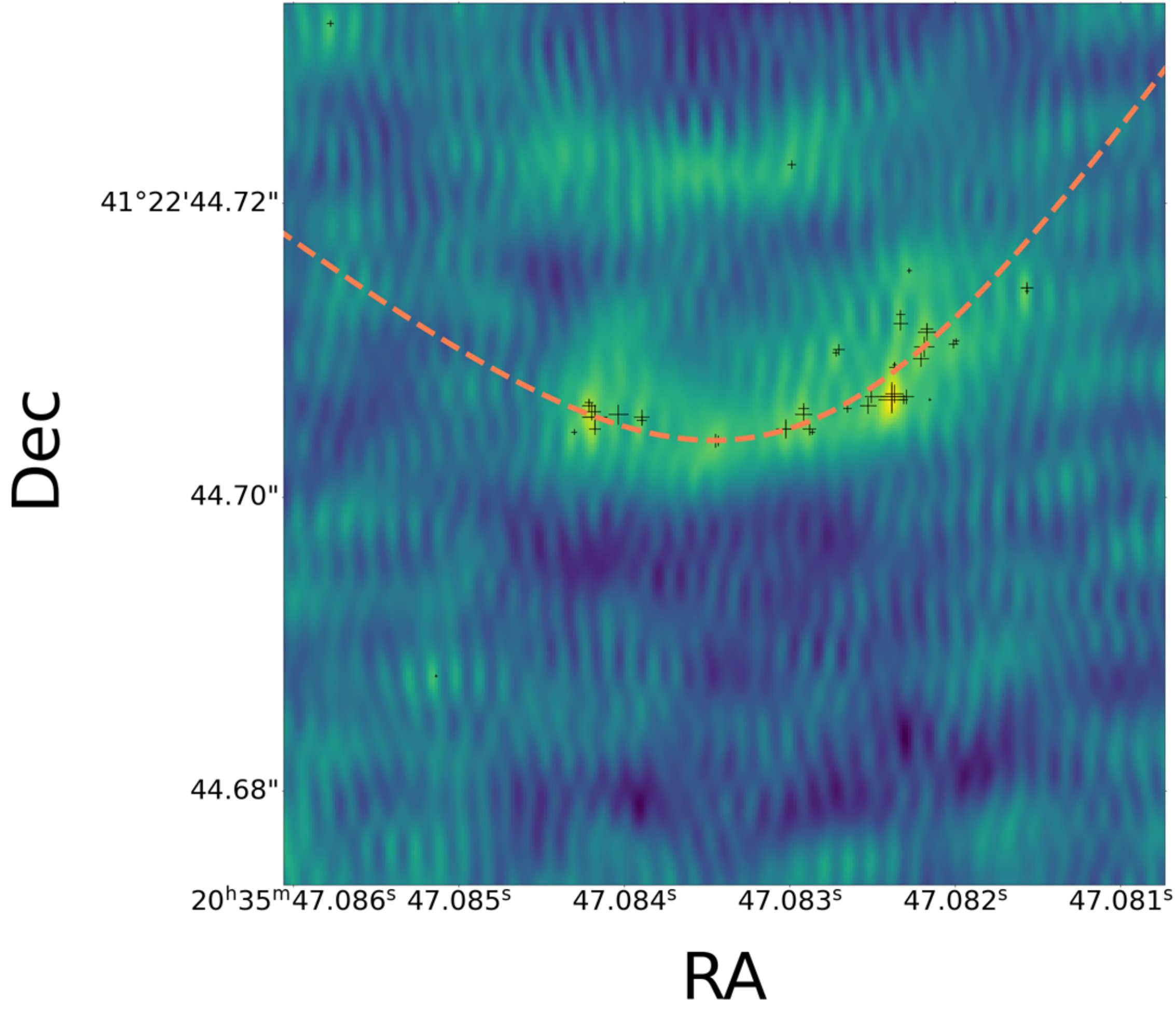} 
    \caption{Clean components and shock cone fittings of WCR images. Black crosses mark the locations of the clean components, with their sizes representing the fitting weights. Orange dashed lines indicate the shape of shock cones. Top panel: fitting result of the BW148a image; bottom panel: result of the EZ025 image.}
    \label{fig:wavefront_fitting}
\end{figure}

Theoretically, the shock front formed by the interaction of isotropic stellar winds should exhibit stable axisymmetry shape about the line connecting the two stars \citep{2003A&A...409..217D}. Thus, we initially assumed that the WCR would have a strictly symmetric shape relative to the binary axis. However, after imaging the VLBI observation data, we found that the WCR's shape is not as symmetric as expected. The turbulence within the stellar winds and variations in u-v sampling across different epochs introduce random distortions into our VLBI images. Therefore, to address how to measure the orientation of a WCR with a randomly distorted unstable shape across different epochs, we develop two methods: 

(I) Simplify the WCR as a shock cone contact discontinuity (CD), then fit the shape of the shock cone whose direction can be treated as the direction of binary alignment \citep{2021MNRAS.501.2478M}. The radio interferometric images are represented as collections of clean components. We use the positions of the clean components to fit the CD. The fluxes of 
clean components are used as the weights for the wavefront fitting.  Figure \ref{fig:wavefront_fitting} present two samples of wavefront fitting.

(II) Another method relies on cross-correlation technique. It can be used to reduce the error in position angle measurements caused by random distortions in the images~\citep{2008MNRAS.386..619C}. To address the measurement error caused by the variability of WCR shape, we introduce the cross-correlation function to measure the offset required for the images to overlap as much as possible, thus calculating the proper motion between images from different epochs.
Cross-correlation technique was originally used for comparing images of the same target at different frequencies within the same epoch. Here, we extend its application to compare the positional changes of the same target in the same frequency band across different epochs. 
Since the estimated orbital period of the binary system exceeds 500 years, and the time span of each series of observations is just about one year, the angular changes caused by the binary orbit over this period can be roughly ignored. 
Therefore, when performing cross-correlation, we only consider the translational component. After determining the displacement, we align the images based on their relative displacement, and then stack them to obtain a composite image representing the sum of this series of observations. In this stacked image, the main structure of the WCR can be highlighted, while the random shape variations in each epoch are diminished. Finally, we fit the wavefront shape of WCR on the stacked image, deriving a position angle as the average position angle for this series of observations. 

\begin{figure}
    \centering    
    \includegraphics[width=0.48\textwidth]{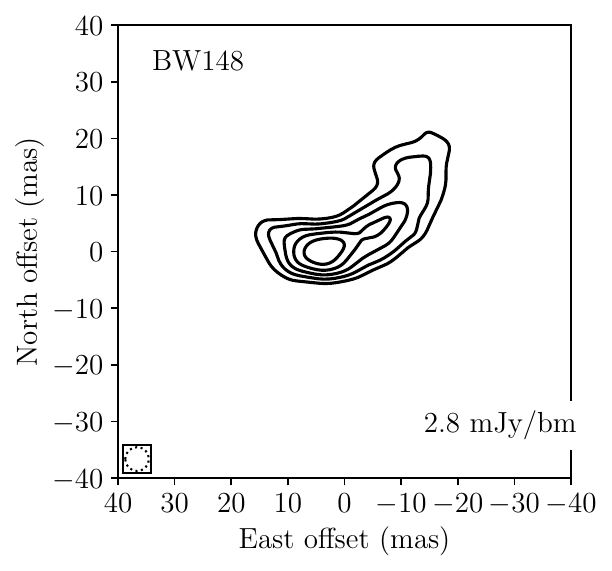}
    \includegraphics[width=0.48\textwidth]{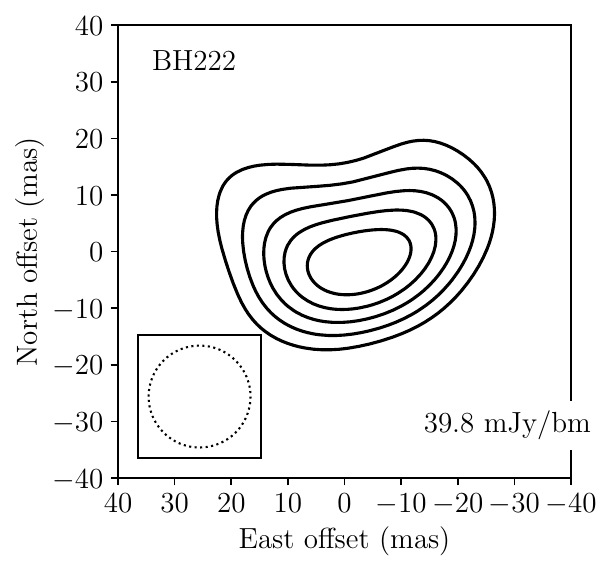}    
    \caption{Stacked images of BW148 and BH222 series. Top panel: Stacked image of BW148 series with restoring beam size 4 mas, $\sigma$ = 0.4 mJy~bm$^{-1}$, and contour level is (3,4,5,6,7) $\times$ 0.4 mJy~bm$^{-1}$; bottom panel: Stacked image of BH222 series with restoring beam size $\sim$20 mas, $\sigma$ = 5 mJy~bm$^{-1}$, and contour level is (3,4,5,6,7) $\times$ 5 mJy~bm$^{-1}$.}
    \label{fig:stacked}
\end{figure} 

For the HST and VLA images of the binary components, we first performed two-dimensional Gaussian fitting to obtain their positions and corresponding uncertainties, using the AIPS task JMFIT.
Furthermore, since the WCR appears as an ellipse in MERLIN observations, we performed two-dimensional Gaussian brightness distribution fitting on the entire 'elliptical' WCR using the AIPS task JMFIT.
We defined the position angle of the binary as the vertical direction of the long axis of the ellipse.
Table~\ref{tab:pa} summarizes the position angles derived from all observed and collected images.

\subsection{Rotation of the binary and WCR} 
\label{sec:rotation} 

Given the differing physical principles underlying the images described in the previous section, we separately analyzed the rotation of the binary and the position angle of the WCR over time to mitigate possible systematic errors. 
To estimate the rotational angular velocity of the binary, we performed linear fitting on the two datasets of binary position angles independently using the least squares method,
with the root-mean-square error as the estimation uncertainties. 

\begin{figure}[H]
    \centering
    \includegraphics[width=0.5\textwidth]{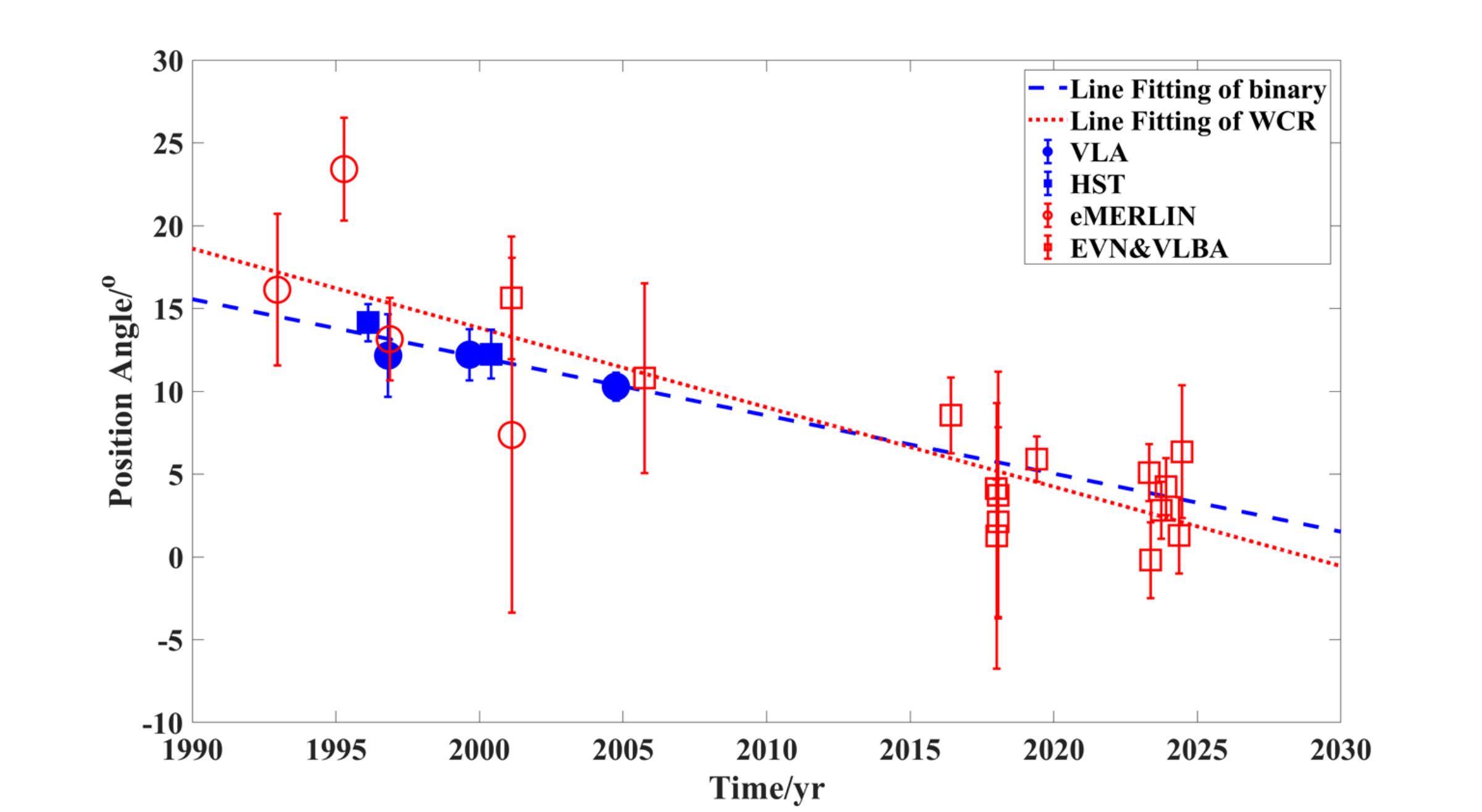}
        \includegraphics[width=0.5\textwidth]{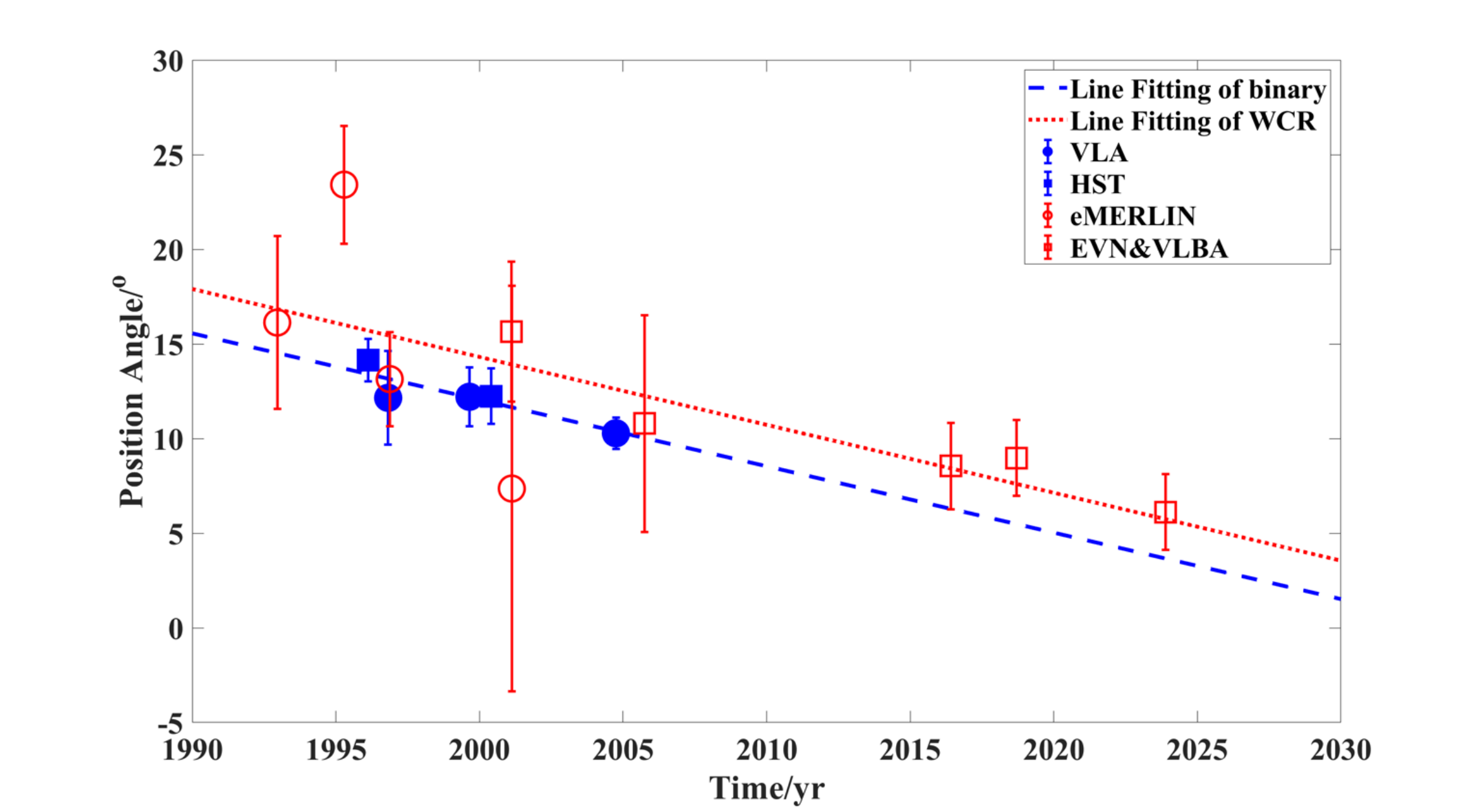}
    \caption{Linear fittings of position angle over time. Four kind of markers represent four different data types, in which the solid dots represent the binary alignment data, and the hollow dots represent the WCR data. The dashed line is the fitting for binary data analysis, and the dotted line is the fitting for WCR data analysis. Top: fitting results of method (I); bottom: fitting results of method (II).
    }
    \label{fig:tdfit}
\end{figure}

For the data-set of binary components, the measured value is 0.351 $\pm$ 0.174 deg yr$^{-1}$. For the WCR, the two methods yield results: 0.479 $\pm$ 0.075 deg yr$^{-1}$ for method (I) and 0.359 $\pm$ 0.148 deg yr$^{-1}$ for method (II).
All these values are consistent within their combined uncertainty.
Treating the binary and WCR data sets as a single entity, we find that the angular velocities are 0.444 $\pm$ 0.056 deg yr$^{-1}$ for method (I) and 0.320 $\pm$ 0.103 deg yr$^{-1}$ for method (II). 
The datasets from both methods, along with their linear fitting results, are shown in Figures~\ref{fig:tdfit} and~\ref{fig:tdfittotal}.
It is noteworthy that, although we initially anticipated a difference—specifically an orbital phase lag—between the fitting lines of the binary and WCR rotations, our observations reveal that this difference is not significant given the current measurement uncertainty. This is particularly true for the larger position angle uncertainties from the WCR compared to those from the binary components. As a result, it remains unclear whether this deviation is real. The potential causes and implications of this deviation will be discussed in Section \ref{sec:discussion}.

\section{Discussion}
\label{sec:discussion}

\subsection{Orbit period}
\label{sec:style}

\begin{figure}[H]
    \centering
    \includegraphics[width=0.5\textwidth]{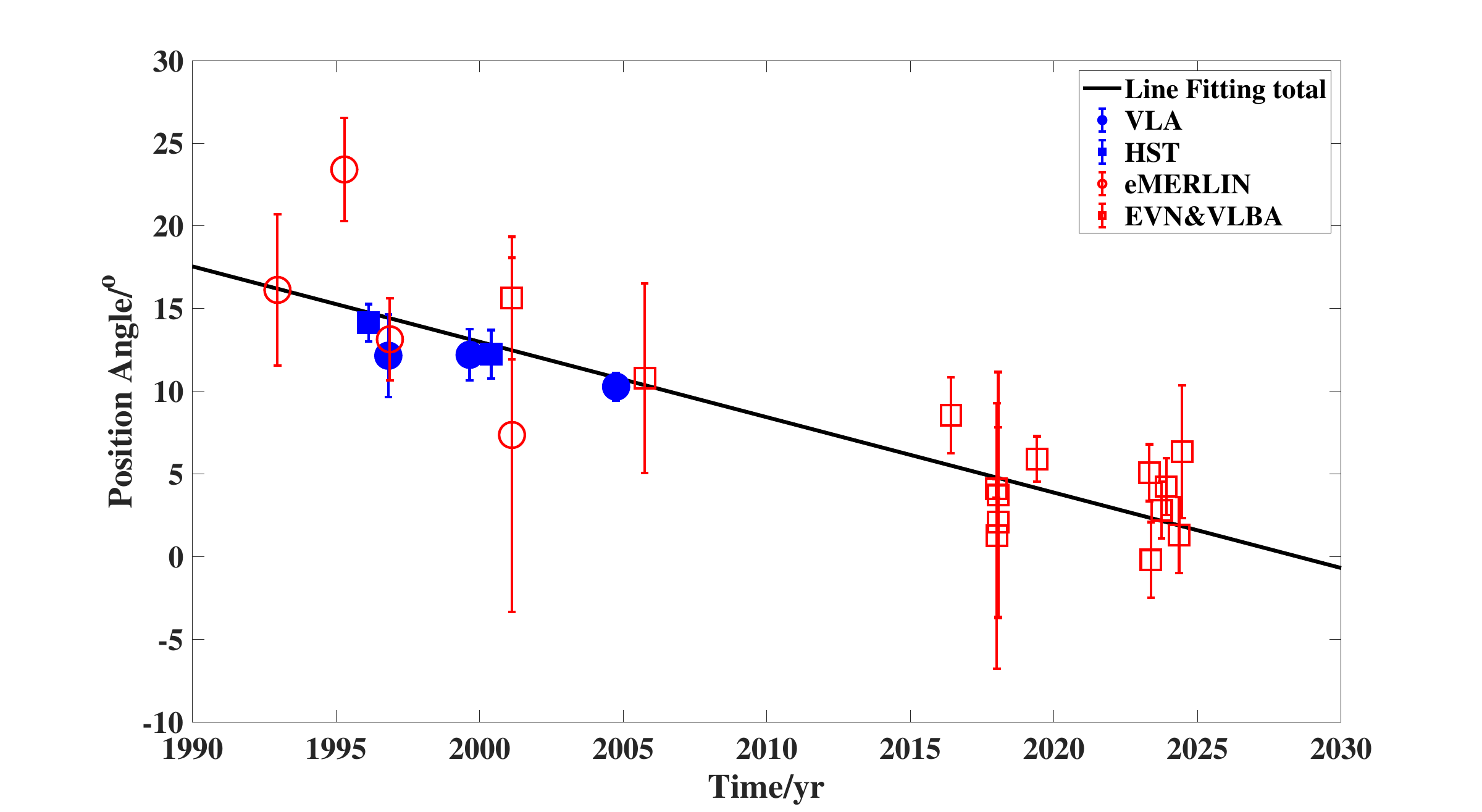}
    \includegraphics[width=0.5\textwidth]{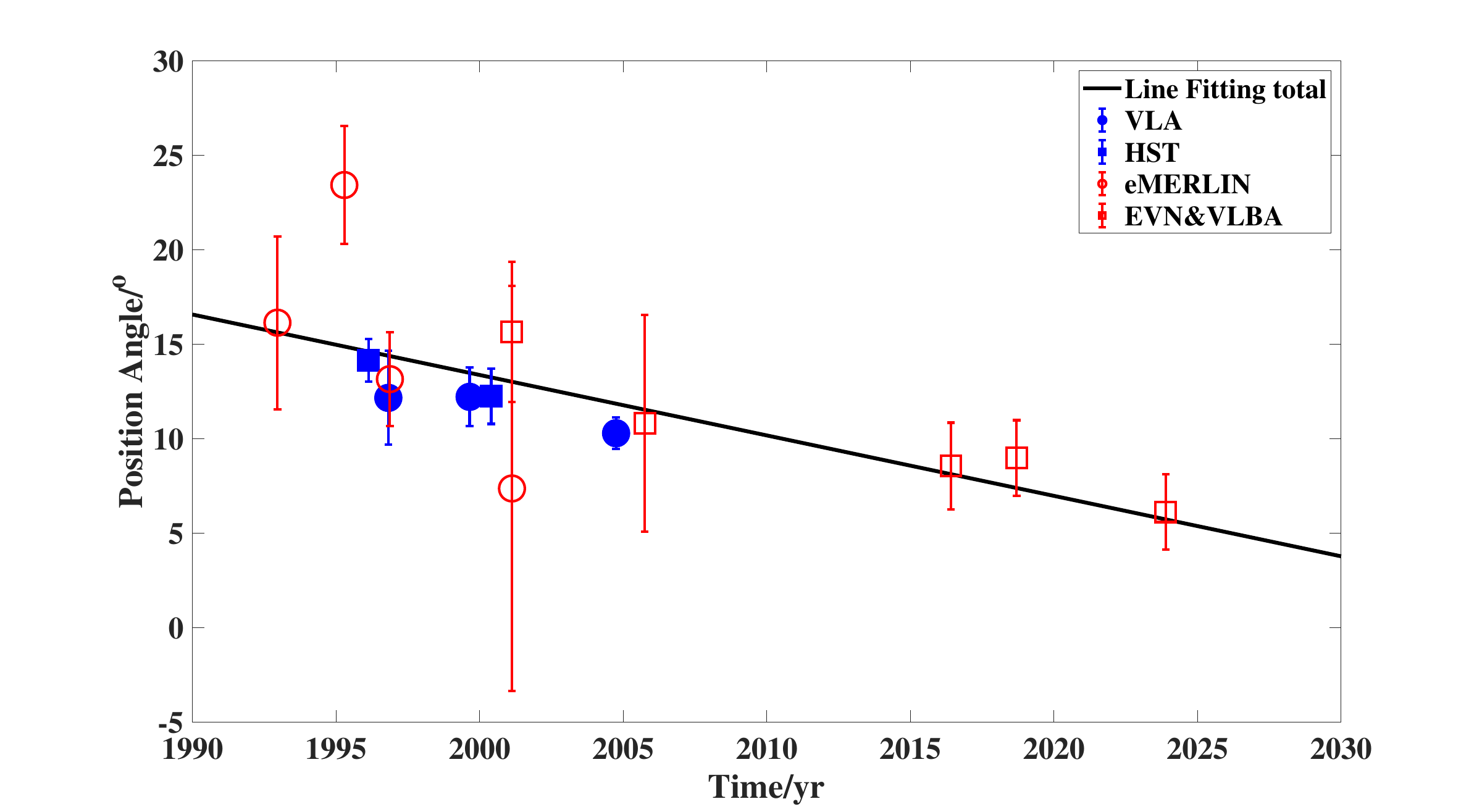}
    \caption{Same as Figure~\ref{fig:tdfit}, but fitting the binary and WCR as a whole sample together. Top panel: the fitting results of method (I); bottom panel: fitting results of method (II).
    } 
    \label{fig:tdfittotal}
\end{figure}

The limited time span of the observational epochs only allows us to cover a small portion of WR 146's orbital period, 
preventing us from capturing the complete orbit.
To minimize the number of free parameters and simplify the discussion, we adopt a face-on circular orbit model to estimate the orbital period.
As shown in Figure~\ref{fig:tdfittotal}, the nearly identical angular velocities of the binary and WCR rotations yield orbital periods of 810$^{+120}_{-90}$ yrs for method (I) and 1120$^{+540}_{-270}$ yrs for method (II) under the face-on circular orbit assumption.

Certainly, the above angular velocity estimation is based on the simplest assumption of a face-on circular orbit. The actual angular velocity of the binary system may require consideration of more complex possibilities, such as:

(I) The binary's orbit may have a significant inclination angle. In this case, the measured angular velocity represents the projection of the orbital motion on the plane of the sky, and the actual angular velocity would be greater than the current measurements;

(II) WR 146 system might be on a high-eccentricity elliptical orbit, with the binary currently near the apastron. In such high-eccentricity cases, the measured angular velocity of the binary would vary significantly, and the current angular velocity estimation may lead to an orbital period longer than the actual orbital period.

Thus, the above estimation should be considered as an upper limit for the binary's orbital period, serving as a reference. More observational data are needed to support and verify this conclusion. Therefore, further observations are essential for more accurate measurements of the binary distance or the angular velocity of this binary system. These data will help determine the orbital eccentricity, offering validation or refutation of this evolutionary hypothesis.

\begin{figure}[H]
    \centering    
    \includegraphics[width=0.3\textwidth]{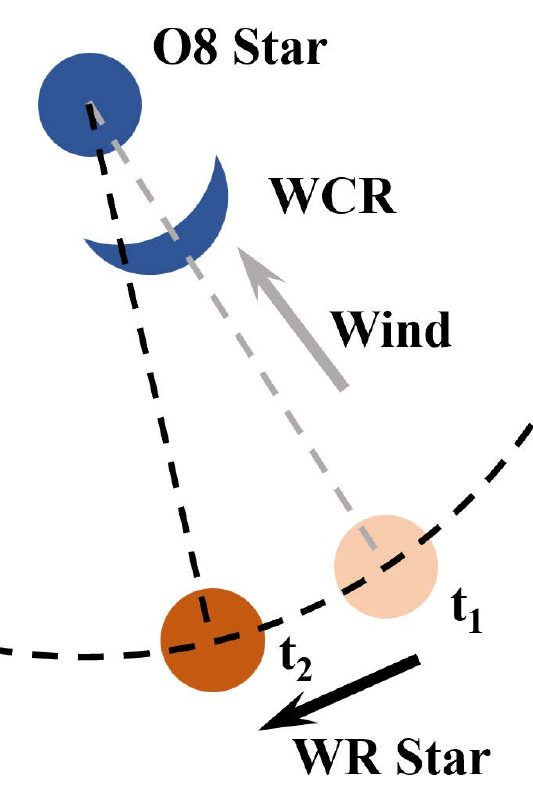}     
    \caption{Diagram of time lag. Stellar wind leaves WR star at $t_1$, but while stellar wind is transferring to WCR, the binary also keeps rotating, when stellar wind reaches WCR at $t_2$, WR star has moved to the location at $t_2$. This results in a discrepancy between the binary rotation and the WCR rotation.}
    \label{fig:td}
\end{figure}   

\subsection{Orbital phase lag between the rotation of the binary components and the WCR}

In Section \ref{sec:rotation}, we separated the data for measuring the position angles of the binary components from the images of the two components and the WCR. Although the orientation of the WCR can be used to derive the position angle, there may be systematic biases between these two datasets.
This may occurs because it takes time for the stellar wind to travel from the star's surface to the WCR. During this travel time, the binary system continues its orbital motion, meaning that the orientation of the WCR at a given moment corresponds to the orientation of the binary components at an earlier time.
This leads to an orbital phase lag between the binary and WCR rotations, as illustrated in Figure \ref{fig:td}. Therefore, our next step is to quantify the extent of the phase lag caused by the travel time of the stellar wind.

\begin{figure*}[htbp]
    \centering    
    \includegraphics[width=0.95\textwidth]{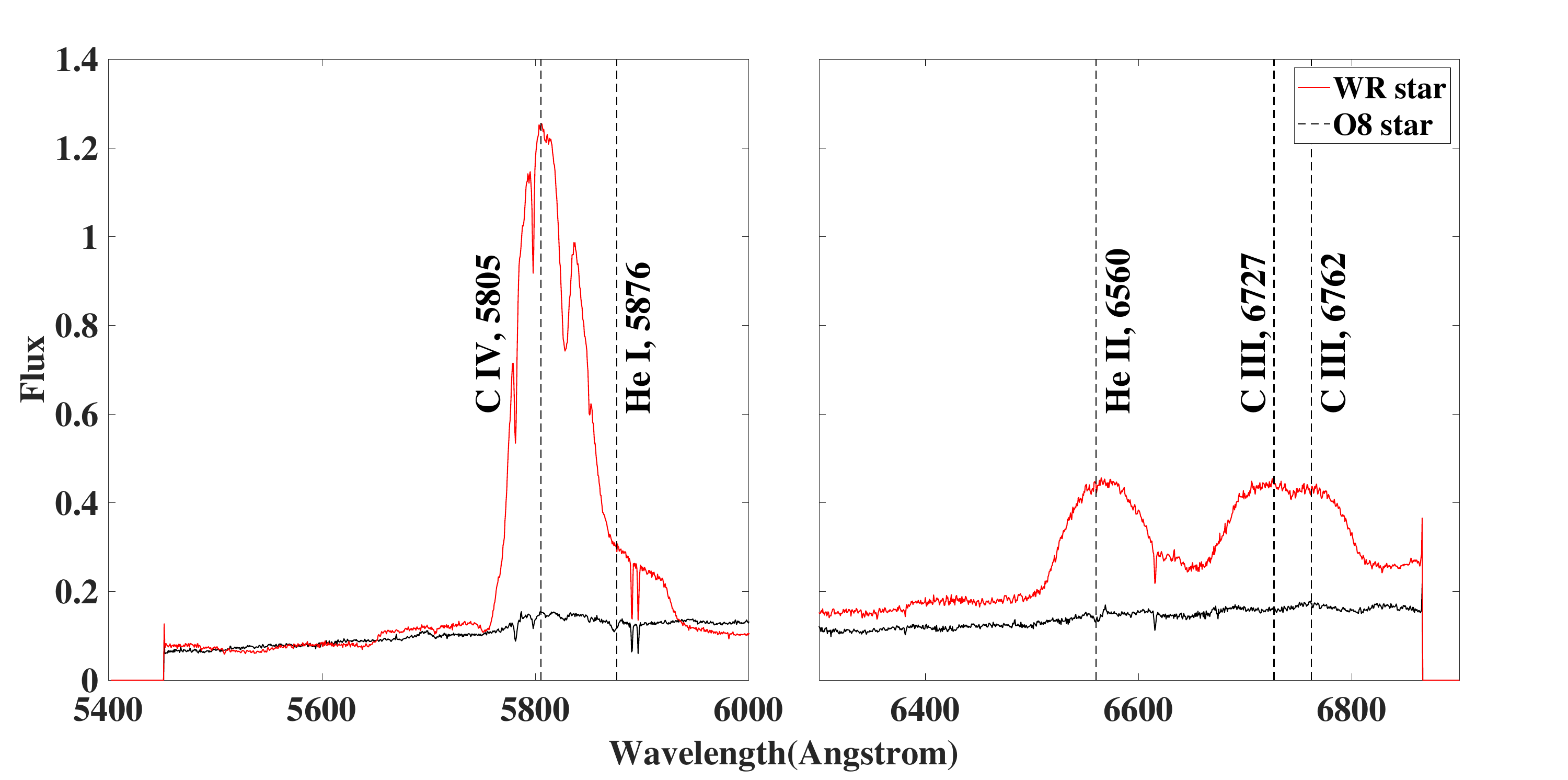} 
    \caption{Binary spectrum observed by HST-STIS in 2000, the binary spectrum can be separated. Wavelength between 5400 and 7000\AA, with a resolution 0.554 \AA$\cdot$pixel$^{-1}$.All noticeable broadlines have been identified by \citet{2001AJ....122.3407L}. The unit of radiation flux is $10^{-12}$erg s$^{-1}$ cm$^{-2}$ \AA$^{-1}$ arcsec$^{-2}$.
    }
    \label{fig:spec}
\end{figure*}

\begin{figure*}[htbp]
    \centering    
    \subfigure[]{
        \begin{minipage}[t]{0.45\linewidth}
            \centering
            \includegraphics[width=7.5cm]{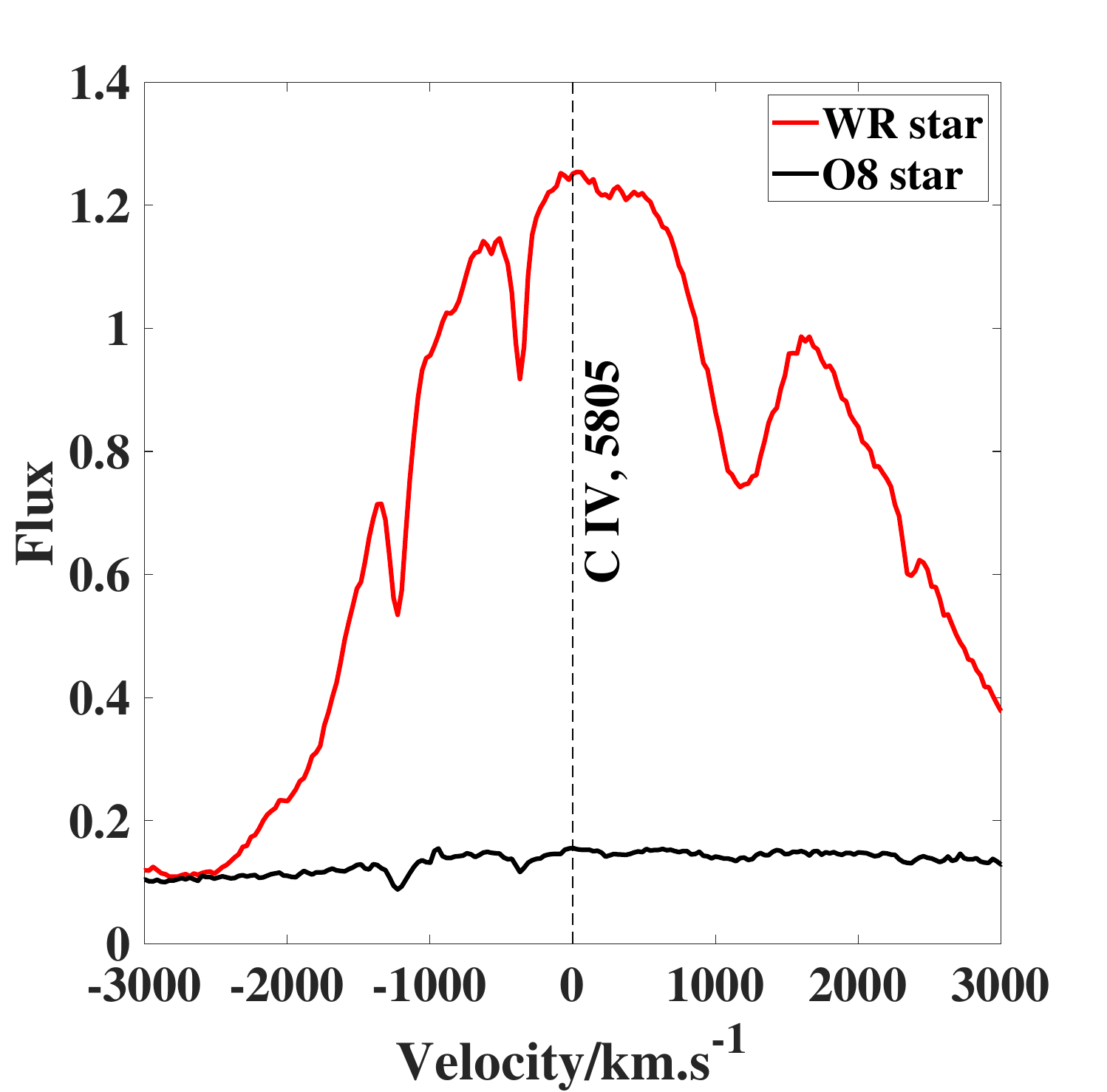}
        \end{minipage}%
    }%
    \subfigure[]{
        \begin{minipage}[t]{0.45\linewidth}
            \centering
            \includegraphics[width=7.5cm]{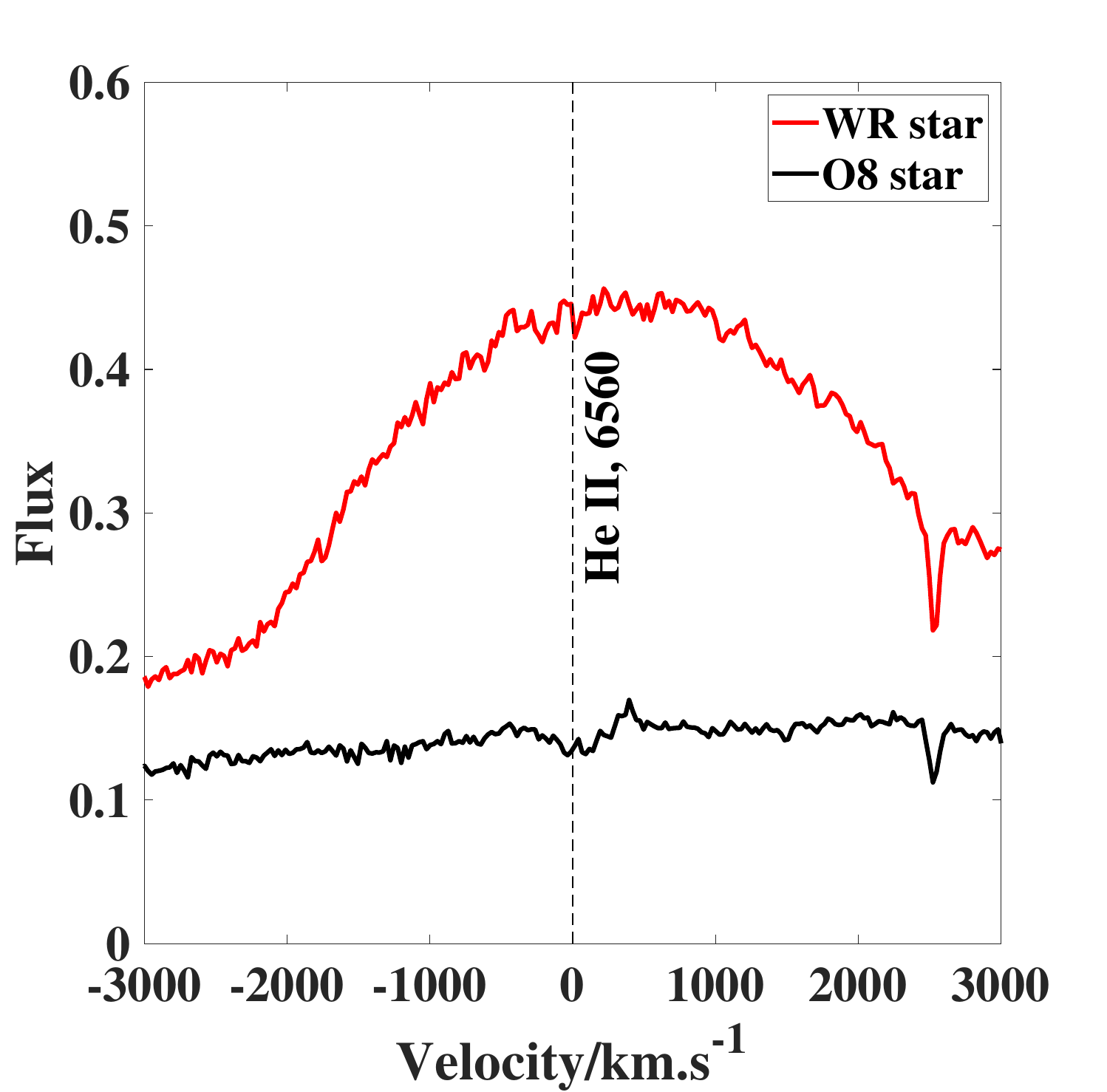}
        \end{minipage}%
    }%
    \quad
    \subfigure[]{
        \begin{minipage}[t]{0.45\linewidth}
            \centering
            \includegraphics[width=7.5cm]{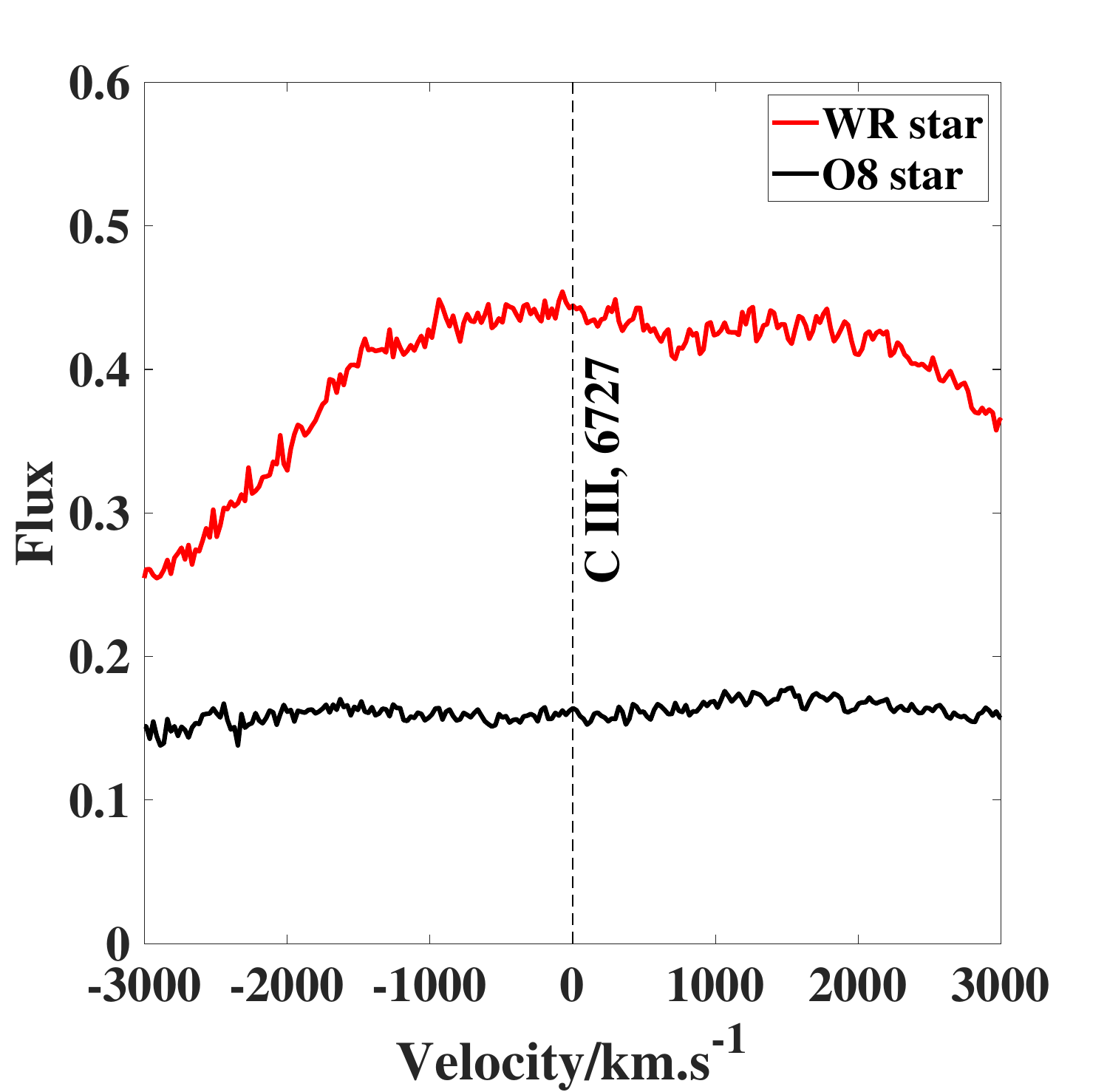}
        \end{minipage}
    }%
    \subfigure[]{
        \begin{minipage}[t]{0.45\linewidth}
            \centering
            \includegraphics[width=7.5cm]{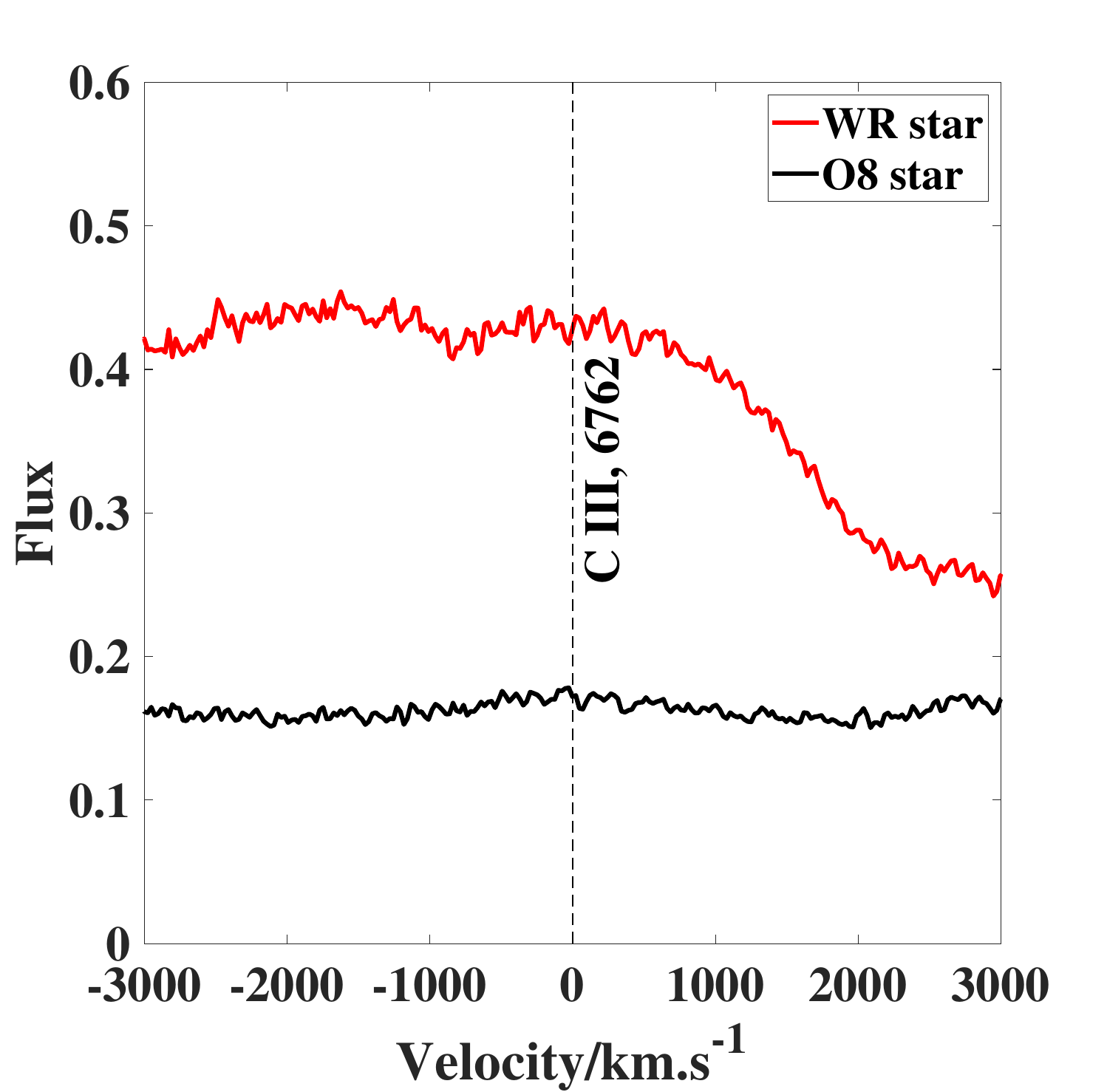}
        \end{minipage}
    }%
    
    \centering
    \caption{Emission lines used for estimating stellar wind velocity in WR 146 spectrum. (a)C IV, $\lambda=5805~\AA$; (b)He II, $\lambda=6560$\AA; (c)C III, $\lambda=6727$\AA; (d)C III, $\lambda=6762$\AA. The spectra display broad line emissions primarily originating from the stellar wind of the WR star, while no discernible line emission was observed from the O8 star. The unit of radiation flux is also $10^{-12}$erg s$^{-1}$ cm$^{-2}$ \AA$^{-1}$ arcsec$^{-2}$.
    }
    \label{fig:lines}
\end{figure*}

To determine the travel time of the stellar wind, we first need to estimate the terminal velocity of the stellar wind. For this, we refer to the spectrum of WR146 observed by HST-STIS). The characteristic broad emission lines from WR stars typically originate from the outflowing wind. Therefore, the widths of these emission line wings can be used to estimate the maximum velocities of the stellar winds~\citep{2007ARA&A..45..177C}. We present the HST spectrum in Figure~\ref{fig:spec} and perform Gaussian fitting on the emission lines shown in Figure~\ref{fig:lines}.
Afterward, we determine the wind velocity by measuring the width of the emission line wings. From the HST spectra,
we calculate the average terminal velocity of the stellar wind from the WR star to be 2,400 $\pm$ 500 km s$^{-1}$. Since the O8 star shows only marginally detected line emission, we adopt the typical terminal velocity for an O8 type star of approximately 1,900 km s$^{-1}$, as reported by \citet{2000ARA&A..38..613K}.

Here, we assume a constant velocity for the stellar winds, with the WR and O8 stars traveling at velocities of 2,400 km s$^{-1}$ and 1,900 km s$^{-1}$, respectively, from the stellar surface to the WCR \citep{2020ApJ...897..135Z}. 
Therefore, adopting a distance of 1.2 kpc and a face-on orbit, the angular separation of 152 mas between the binary components corresponds to a physical separation:

\begin{equation}
  R=(152\pm2)~{\rm mas}\times1.2^{+1.0}_{-0.4}~{\rm kpc}=182.4^{+156.4}_{-62.4}~{\rm au} 
  \label{eq:BinarySeparation}
\end{equation}

The propagation time of the stellar wind from the WR star to the WCR is 0.38$^{+0.33}_{-0.13}$  years, while it takes 9.1$^{+7.6}_{-3.0}$ days for the wind to travel from the O8 star to the WCR. 
Therefore, both the 0.38 years and 9.1 days phase lag are much smaller than the binary orbital period of approximately 1000 years, we conclude that the orbital phase lag is not particularly significant.

\subsection{Masses of binary stars}

Total Mass of binary stars can be estimated according to Kepler's Third Law, 
\begin{equation}
T^2=\frac{4\pi^2}{GM}R^3
\label{e:Kepler}
\end{equation}

where $M$ is the total mass of the binary, $T$ is the orbital period, and $R$ is the binary separation. 
In addition to the distance of 1.2$^{+1.0}_{-0.4}$ released by Gaia DR2, another perspective suggests that WR 146 may be a member of the Cyg OB2 association \citep{1991AJ....101.1408M}, which is located at a distance of 1.5 - 2.0 kpc, as suggested by \citet{2021MNRAS.508.2370Q} 
Using a distance of \( 1.2^{+1.0}_{-0.4} \) kpc, along with the observed separation of 152 mas between the binary components and an estimated orbital period of 810 - 1,120 years, the results suggest that the total mass of the binary system is approximately 1.45 - 57.62 M$_\odot$. While with a distance range of 1.5 - 2.0 kpc, the total mass of the binary system is estimated to be between 9.55 - 43.29 M$_\odot$.

Next, to estimate the mass of each component, we compare the line widths in the HST spectrum of the WR star with Potsdam Wolf-Rayet (\textit{PoWR}) atmosphere models ~\footnote{\url{https://www.astro.physik.uni-potsdam.de/PoWR/}} \citep{2003A&A...410..993H, 2012A&A...540A.144S}, to identify the best match for our WR star. The \textit{PoWR} model considers non-local thermal equilibrium, spherical expansion, and metal line blanketing. It uses the line shapes of typically emission lines of WR stars (such as He, C, N, and O lines) as the basis for matching the stellar parameters of the target star. For Galactic WR stars, PoWR categorizes them into two groups: WC and WN. WR star in the WR 146 system, is classified as a WC6 star \citep{2001AJ....122.3407L}.

\begin{figure*}
    \centering    
    \includegraphics[width=0.9\textwidth]{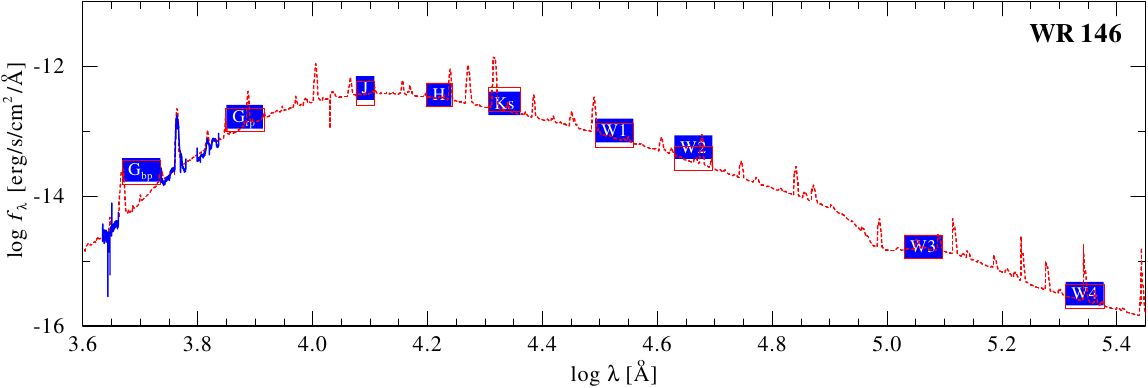} 
    \caption{Spectral energy distribution of WR 146: Observed photometry (blue boxes) from Gaia EDR 3, 2MASS, and WISE and flux-calibrated spectra from HST STIS (blue lines), plotted together with the emergent spectrum of the best-fit \textit{PoWR} model (red dashed line). For a better comparison also the resulting photometric values from the \textit{PoWR} model are shown as red boxes.}
    \label{fig:PoWR_sed}
\end{figure*}

\begin{figure*}
    \centering    
    \includegraphics[width=0.9\textwidth]{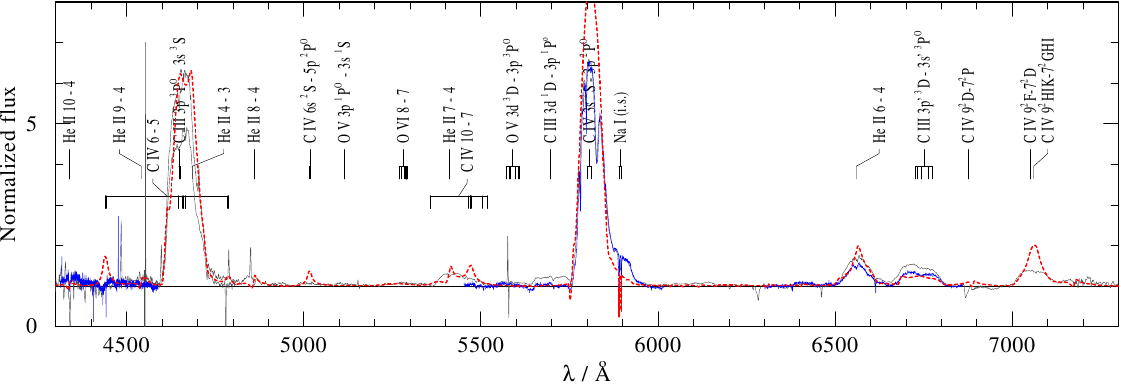} 
    \caption{Details of the optical spectrum of WR 146: HST-STIS observations (blue lines) and observations from Calar Alto Observatory in 1991 (thin gray lines) \citep{1995A&AS..113..459H}, both normalized by eye, shown together with the continuum-normalized spectrum of the best-fit \textit{PoWR} model (red dashed line). Most important spectral lines are labeled.}
    \label{fig:PoWR_opt}
\end{figure*}

To infer the basic parameters of WR 146 we calculated a series of stellar atmosphere models with \textit{PoWR} and compared them with observed spectra and photometry. Given the complexity of WR atmospheres, achieving a perfect match for all optical emission lines simultaneously is inherently challenging: for example, the models cannot reproduce the round-line profile of He II 6560 \AA  and C IV 7050/61 \AA \citep{2014A&A...562A.118S}. Nevertheless -- given the limited observational material and the contribution by the O8 star in the spectra -- our best-fit model reproduces most of the spectral features originating from helium, carbon, and oxygen emission lines (see Figure~\ref{fig:PoWR_sed}). Using the distance estimate of 1.2$^{+1.0}_{-0.4}$ kpc from \citet{2018AJ....156...58B}, the fit of the synthetic SED to the observed photometry (see Figure~\ref{fig:PoWR_opt}) implies a luminosity of the WR star of log (L/L$_{\odot}$) =6.14$^{+0.50}_{-0.37}$. Based on this result, we calculate the stellar mass according to the mass-luminosity relation by \citet{1989A&A...210...93L} for massive WC stars and obtain a stellar mass of 40$^{+82}_{-18}$ M$_{\odot}$. The parameters of our best-fit model are summarized in Tab~\ref{tab:PoWR}.

The typical mass of an O8-type main-sequence star is approximately 15 to 30 M$_{\odot}$ \citep{2003A&A...404..975M}. Consequently, the total mass of the binary system is expected to range from 37 to 152 M$_{\odot}$. This range overlaps with the previously mentioned values of 1.45 M$_{\odot}$ to 57.62 M$_{\odot}$, reflecting the significant uncertainties in current estimates for the period and distance. To achieve more accurate estimates of the physical parameters of the WR 146 system, including the masses of its components, accurate distance measurements are essential.

\section{Summary}
\label{sec:sum}

In this paper, we present our VLBI observations of the CWB system WR 146 using the EVN and VLBA, aimed at extending the time range for more accurately determining its orbital period. Combined with archival radio and optical observations from the EVN, VLBA, VLA, eMERLIN, and HST, our study encompasses 24 effective epochs from 1993 to 2024. We summarize our results as follows:

\begin{itemize} 

\item[(1)] We categorized our samples into two groups: WCR measurements and binary measurements. Linear regression was performed for each group separately. For the rotation of the WCR, we explored two measurement methods: (I) fitting the shock cone and (II) employing a cross-correlation technique. Based on these methods, we determined the rotation angular velocity for the binary to be 0.3510 $\pm$ 0.1744 deg yr$^{-1}$. For the WCR, we measured angular velocities of 0.4792 $\pm$ 0.0750 deg yr$^{-1}$ using method (I) and 0.3588 $\pm$ 0.1483 deg yr$^{-1}$ using method (II). When combining all data points, the overall angular velocities were found to be 0.4436 $\pm$ 0.0565 deg yr$^{-1}$ for method (I) and 0.3198 $\pm$ 0.1033 deg yr$^{-1}$ for method (II). Based on these results, we estimate the binary orbital period to be 810$^{+120}_{-90}$ years for method (I) and 1120$^{+540}_{-270}$ years for method (II). Therefore, we conclude that the orbital period of the binary is approximately 1,000 years.

\item[(2)] 
We explored a potential phase lag between the binary star and the WCR, which may arise from the time it takes for the stellar wind to travel from the WR/O8 star to the WCR. This phase lag is estimated to be approximately 0.38$^{+0.33}_{-0.13}$ years. However, no significant translational motion has been observed in either the binary or the WCR rotation, indicating that the phase lag phenomenon is not particularly pronounced.

\item[(3)]
Based on the calculations above, we conducted a preliminary estimate of the binary system's mass. The results reveal a wide range of estimated masses, varying from significantly smaller to comparable with typical massive binaries. We attribute this variability to inaccuracies in distance measurements, which contribute to the large uncertainty in the estimated mass of the binary system.

\end{itemize} 

Although our determined orbital parameters still exhibit relatively large uncertainties, they highlight the types of observations needed in the future to refine these parameters. 

The accuracy of observations is expected to improve significantly with the development of high-sensitivity instruments for radio interferometry, such as the Square Kilometer Array (SKA). Additional gains in resolution can be achieved by integrating SKA with existing VLBI arrays. In the optical domain, advancements in observational instruments and techniques, including Gaia for parallax measurements and the James Webb Space Telescope (JWST) for spectroscopy, will enable more precise determinations of distances and radial velocities. These improved measurements will provide more accurate parallax data and contribute to a better understanding of the physical processes in the WCR.

\section*{Acknowledgments} 
\label{sec:style}

We appreciate the anonymous referee for helpful comments that improved this manuscript. We appreciate Gabor Orosz, Sandor Frey, Zhiqiang Shen, Junzhi Wang, Wei Zhang, Hong Wu and Qingkang Li, for their valuable discussions throughout this research. This work is supported by the National Natural Science Foundation of China (NSFC) under grant Nos. U2031212. This work is also supported by the CAS 'Light of West China' Program (grant No. 2021-XBQNXZ-005), the National SKA Program of China (grant No. 2022SKA0120102), and the Tianshan Talent Training Program (grant No. 2023TSYCCX0099). Xiaofeng Li was supported by the National Natural Science Foundation of China (No. 12203014). The HST data presented in this article were obtained from the Mikulski Archive for Space Telescopes (MAST) at the Space Telescope Science Institute. The specific observations analyzed can be accessed via dataset[doi:10.17909/pg5j-ta08]https://doi.org/10.17909/pg5j-ta08.

\bibliography{ref}
\bibliographystyle{aasjournal}

\begin{table}
\centering
\caption{Datasets for images of resolved binary components.} 
\setlength\tabcolsep{3pt} 
\begin{tabular}{cccccc} 
\hline
\hline 
\begin{tabular}[c]{@{}l@{}}Program Code\\ \end{tabular} & 
\begin{tabular}[c]{@{}l@{}}Epoch        \\ \end{tabular} & 
\begin{tabular}[c]{@{}l@{}}Telescope$\&$Instrument    \\ \end{tabular} & 
\begin{tabular}[c]{@{}l@{}}Band          \\\end{tabular}     & 
\begin{tabular}[c]{@{}l@{}}Resolution    \\(mas) \end{tabular} &  
\begin{tabular}[c]{@{}l@{}}Observed Time \\  \end{tabular}   \\
\hline 
AD391     & 1996-OCT-26   & VLA-A     & K       & $105\times 84$  & 120 min \\
AD426     & 1999-AUG-26   & VLA-A     & K       & $106\times 79$  & 100 min \\
AD502     & 2004-OCT-01   & VLA-A     & Q       & $39 \times 27$  &  20 min \\
\hline
u33u19    & 1996-FEB-17   & HST-WFPC2 & F555W   & 50              &  1.6 sec \\
o5d502    & 2000-MAY-26   & HST-STIS  & MIRVIS  & 50      &  7.7 sec \\
\hline
\end{tabular} 
\tablecomments{
      The archival data of binary system WR~146 observed by VLA and HST. Column 1
    present the data program code, column 2 present the observing date of archival data, columns 3-6 give the telescope, 
    observing waveband, angular resolution, and
    on source or exposure time, respectively. VLA-A denotes VLA in A configuration, 
    WFPC2 denotes Wide Field Planetary Camera 2, and STIS denotes Space Telescope Imaging Spectrograph,
    F555W is central wavelength of imaging waveband is 555 nm.
}
      
\label{tab:VLAHST}     
\end{table} 

\begin{table}
\centering
\caption{Radio inteferometer observations of WR 146.} 
\setlength\tabcolsep{3pt} 
\begin{tabular}{cccccc} 
\hline
\hline 
\begin{tabular}[c]{@{}l@{}}Array \end{tabular} & 
\begin{tabular}[c]{@{}l@{}}Program\\ Code \end{tabular} & 
\begin{tabular}[c]{@{}l@{}}Epoch\end{tabular} & 
\begin{tabular}[c]{@{}l@{}}Band \end{tabular}     & 
\begin{tabular}[c]{@{}l@{}}Resolution \\(mas) \end{tabular} &  
\begin{tabular}[c]{@{}l@{}}Observed \\Time (min) \end{tabular} \\ 
\hline 
EVN     & EZ025     & 2016-MAY-30     &C       & $5.2\times 4.5$   & 480 \\
        & ED017a    & 2001-FEB-12     &C       & $11 \times 8.4$   & 180  \\
\hline
VLBA& BD111      & 2005-NOV-26      &C     & $7.8 \times 6.1$  & 540 \\
    & BW148a     & 2023-APR-27      &C     & $3.6 \times 2.4$  & 480 \\
    & BW148b     & 2023-MAY-18      &C     & $3.9 \times 2.2$  & 480 \\
    & BW148c     & 2023-SEP-29      &C     & $3.6 \times 1.7$  & 480 \\
    & BW148d     & 2023-NOV-29      &C     & $3.0 \times 1.6$  & 480 \\
    & BW148e     & 2023-MAY-09      &C     & $3.3 \times 1.4$  & 480 \\
    & BW148f     & 2023-JUN-16      &C     & $4.1 \times 3.5$  & 480 \\
    & BH222a     & 2017-DEC-29      &L     & $20 \times 15$    & 240 \\
    & BH222b     & 2018-JAN-04      &L     & $19 \times 11$    & 240 \\
    & BH222c     & 2018-JAN-21      &L     & $18 \times 11$    & 240 \\
    & BH222d     & 2018-JAN-23      &L     & $18 \times 12$    & 240 \\
    & BH222e     & 2019-MAR-24      &L     & $29 \times 13$    & 240 \\
    & BH222f     & 2019-MAY-30      &L     & $22 \times 12$    & 240 \\
\hline
 MERLIN & 92DECA4993 & 1992-DEC-01  &C            & $50\times 50$ & 275 \\
        & 95APRD4994 & 1995-APR-15  &C            & $50\times 50$ & 770 \\
        & 96NOVA4994 & 1996-NOV-01  &C            & $40\times 50$ & 640 \\
        & 01FEBE4990 & 2001-FEB-10  &C            & $50\times 50$ & 300 \\
\hline
\end{tabular} 
\tablecomments{
      This table appear gathered data observed by several radio
      interferometry arrays. Column 1 is observed telescope; column 2
      is data program code; column 3 is epoch; column 4 is
      waveband; column 5 is estimated data resolution and column 6 is the integration time.
      }      
\label{tab:radio}     
\end{table}

\begin{table}[H] 
\centering
\caption{Position angle of binary measured by WR146 binary and WCR images.} 
\setlength\tabcolsep{0pt} 
\begin{tabular}{ccccc} 
\hline 
\begin{tabular}[c]{@{}l@{}}Object \end{tabular} & 
\begin{tabular}[c]{@{}l@{}}Telescope \end{tabular} & 
\begin{tabular}[c]{@{}l@{}}Programe Code \end{tabular} & 
\begin{tabular}[c]{@{}l@{}}Epoch \end{tabular}     & 
\begin{tabular}[c]{@{}l@{}}Position Angle/$^{\circ}$ \end{tabular} \\
\hline 
WCR     & EVN       & EZ025     & 2016-MAY-30       & 8.55$\pm$2.29 \\
        &           & ED017a    & 2001-FEB-12       & 15.64$\pm$3.70 \\
        \hline 
        & VLBA      & BD111     & 2005-NOV-27       & 10.80$\pm$5.73 \\
        &           & BH222a    & 2017-DEC-29       & 4.12$\pm$0.57\\
        &           & BH222b    & 2018-JAN-04       & 1.26$\pm$8.02\\
        &           & BH222c    & 2018-JAN-21       & 3.72$\pm$7.45\\
        &           & BH222d    & 2018-JAN-23       & 2.10$\pm$5.13\\
        &           & BH222e    & 2019-MAR-24       & Failed fitting\\
        &           & BH222f    & 2019-MAY-30       & 5.90$\pm$1.37\\
        &           & BH222$^{*}$  & 2016-SEP-13    & 8.97$\pm$3.51\\
        &           & BW148a    & 2023-APR-27       & 5.08$\pm$1.72\\
        &           & BW148b    & 2023-MAY-18       & 2.81$\pm$1.72\\
        &           & BW148c    & 2023-SEP-29       & 4.24$\pm$1.72\\
        &           & BW148d    & 2023-NOV-29       & 1.30$\pm$2.29\\
        &           & BW148e    & 2024-MAY-09       & -0.20$\pm$2.29\\
        &           & BW148f    & 2024-JUN-16       & 6.35$\pm$4.01\\
        &           & BW148$^{*}$  & 2023-NOV-22    & 6.12$\pm$2.26\\
\hline
        & MERLIN    & 92DECA4993 & 1992-DEC-01       & 16.1$\pm$4.6 \\
        &           & 95APRD4994 & 1995-APR-15       & 23.4$\pm$3.1 \\
        &           & 96NOVA4994 & 1996-NOV-01       & 13.1$\pm$2.5 \\
        &           & 01FEBE4990 & 2001-FEB-10       & 7.3$\pm$10.7 \\
\hline
\hline
Binary  & VLA       & AD391      & 1996-OCT-26      & 12.1$\pm$2.5 \\
        &           & AD426      & 1999-AUG-26      & 12.2$\pm$1.6 \\
        &           & AD502      & 2004-OCT-01      & 10.3$\pm$0.8 \\
\hline
        & HST       & o5d502     & 1996-FEB-17      & 14.1$\pm$1.1 \\
        &           & u33u19     & 2000-MAY-26      & 12.2$\pm$1.5 \\
\hline
\end{tabular} 
\label{tab:pa} 
~\\ 
~\\ 
\edit1{Notes: } 
Orientation of binaries or WCRs at each epoch. The radio band was obtained by AIPS task JMFIT, and the optical band by DS9. The position angle of the binary observed by VLA and HST and the position angle of the WCR observed by EVN, VLBA and eMERLIN. Raws marked by $^{*}$ are measured by stacked images.
\end{table} 

\begin{table}[H] 
\centering
\caption{Parameters of our best-fit \textit{PoWR} model for WR star.} 
\setlength\tabcolsep{0pt} 
\begin{tabular}{ccc}
\hline
Parameter & value & comment\\
\hline
$E_{B-V}$ [mag] & 2.83 & \\
$T_*$     [kK]          & 63 \\
$\log L$  [$L_\odot$]   &  6.14 & for $d=1.2\,$kpc \\
$R_*$     [$R_\odot$]   & 9.86 & defined at $\tau_{\mathrm Ross}=20$ \\
$R_{\mathrm t}$ [$R_\odot$]  & 10.7 & transformed radius\\
$\dot{M}$ [$M_\odot\,{\mathrm yr}^{-1}$] & $3.3\times 10^{-5}$ \\
$v_\infty$ [km/s] & 3000  & terminal velocity \\
$v_{\mathrm Dop}$ [km/s] & 100 & microtubulence velocity \\
$D$ & 10 & density contrast \\[0.2cm]
$X_{\mathrm He}$ & 0.60 & mass fraction, not fitted \\
$X_{\mathrm C}$  & 0.40 & mass fraction, not fitted \\
$X_{\mathrm O}$  & 0.005 & mass fraction \\
$X_{\mathrm Fe}$ & 0.0016 & mass fraction, not fitted \\
\hline
\end{tabular}

\label{tab:PoWR} 
~\\ 
~\\ 

\end{table} 

\end{document}